\documentclass[preprintnumbers,11pt,a4paper,oneside]{article}
\pdfoutput=1

\usepackage{contour}
\usepackage{xcolor}

\usepackage{mathrsfs}

\contourlength{0.05em}

\usepackage{jheppub}
\usepackage{color}
\usepackage{graphicx}
\usepackage{wrapfig,enumerate,slashed}

\usepackage{footmisc}
\usepackage{amsmath}
\usepackage{wasysym} 
\usepackage{graphicx}
\usepackage{subcaption}
\usepackage{color}
\usepackage{comment}
\usepackage{appendix}
\usepackage{slashed}

\newcommand{\be}{\begin{equation}}
\newcommand{\ee}{\end{equation}}
\newcommand{\beq}{\begin{equation}}
\newcommand{\eeq}{\end{equation}}

\newcommand{\bee}{\begin{eqnarray}}
\newcommand{\eee}{\end{eqnarray}}
\newcommand{\beeq}{\begin{equation}}
\newcommand{\eeeq}{\end{equation}}

\newcommand{\nth}[1]{$#1$\,th}
\newcommand{\bra}[1]{\langle#1 |}
\newcommand{\ket}[1]{|#1 \rangle}
\newcommand{\braket}[2]{\left \langle #1 | #2 \right \rangle}

\newcommand{\sandwich}[3]{\left \langle #1 | #2 | #3 \right\rangle}

\usepackage{tikz}
\usepackage{orcidlink}

\makeatletter
\gdef\@fpheader{}
\makeatother
\begin{document}

\title{Simulating quantum field theories on continuous-variable quantum computers}

\author[a,b]{Steven~Abel\,\orcidlink{0000-0003-1213-907X}}
\author[a]{Michael Spannowsky\,\orcidlink{0000-0002-8362-0576}}
\author[a]{Simon Williams\,\orcidlink{0000-0001-8540-0780}}

\emailAdd{steve.abel@durham.ac.uk}
\emailAdd{michael.spannowsky@durham.ac.uk}
\emailAdd{simon.j.williams@durham.ac.uk}

\affiliation[a]{\vspace{0.1cm} Institute for Particle Physics Phenomenology, Durham University, Durham DH1 3LE, UK}
\affiliation[b]{Department of Mathematical Sciences, Durham University, Durham DH1 3LE, UK}

\preprintA{IPPP/24/10}

\abstract{We delve into the use of photonic quantum computing to simulate quantum mechanics and extend its application towards quantum field theory. We develop and prove a method that leverages this form of Continuous-Variable Quantum Computing (CVQC) to reproduce the time evolution of quantum-mechanical states under arbitrary Hamiltonians, and we demonstrate the method's remarkable efficacy with various potentials. Our method centres on constructing an {\it evolver-state}, a specially prepared quantum state that induces the desired time-evolution on the target state. This is achieved by introducing a non-Gaussian operation using a measurement-based quantum computing approach, enhanced by machine learning. Furthermore, we propose a framework in which these methods can be extended to encode field theories in CVQC without discretising the field values, thus preserving the continuous nature of the fields. This opens new avenues for quantum computing applications in quantum field theory.}

\maketitle


\section{Introduction}

Harnessing the intricate dynamics of quantum mechanics to improve our understanding of fundamental physics has led to the pursuit of computational paradigms that go beyond classical boundaries and allow for complex systems to be simulated with devices which themselves are inherently quantum mechanical. Such devices would leverage the properties of quantum systems, such as superposition and entanglement, to perform calculations directly using the intrinsic dynamics of the system. Two main paradigms have been identified: {\it digital} and {\it analogue} quantum computing. The former encodes information onto systems with a finite number of discrete degrees of freedom, such as the qubit, a two-state quantum system. The latter instead encodes information on systems which are described by operators which have a continuous spectrum. This paradigm is often called Continuous-Variable Quantum Computing (CVQC)~\cite{Lloyd1999, RevModPhys.77.513,Ladd2010,Adesso_2014}. Among the various kinds of CVQC, quantum optics emerges as a fascinating framework, using the infinite-dimensional landscape of photon states to encode and manipulate information~\cite{PhysRevA.73.032318, Kok2007, Slussarenko2019, Killoran_2019, Bromley_2020, Eaton_2022, Taballione2023modeuniversal}. The eigenstates of these operators form an infinite-dimensional Hilbert space, with the continuous-variable analog of the qubit being the so-called \textit{qumode}. 

CVQC, rooted in the continuous spectra of quantum operators, offers various possibilities for simulating the dynamics of quantum particles and fields. By employing such qumodes --- quantum analogues of classical harmonic oscillators --- as the fundamental information units, CVQC provides a natural framework for encoding quantum states in the continuous observables of photons, e.g. their position or momentum. This paradigm allows for implementing Gaussian gate operations, which manipulate the quantum states through transformations that preserve their Gaussian character, thereby enabling a broad range of quantum simulations. However, the true power of CVQC unfolds with the inclusion of non-Gaussian operations, which introduce higher-order interactions essential for achieving universal quantum computation~\cite{Lloyd1999,PhysRevA.100.012326,PhysRevA.100.052301}. These operations, albeit challenging to implement due to the weakly interacting nature of photons, open the door to simulating complex quantum systems with high fidelity. Most existing proposals to simulate quantum systems on CVQC rely on specific ways to induce non-Gaussian effects, such as the Kerr effect. However, the non-Gaussian characteristics introduced by current non-linear optical materials are very weak~\cite{RevModPhys.77.513, Slussarenko2019,Kok2007, Ladd2010, PhysRevA.100.052301}, and constructing an arbitrary non-Gaussian operation is difficult. An alternative approach is achieved by integrating measurement-based quantum computing techniques~\cite{PhysRevLett.86.5188, Booth:2021hvw} and leveraging the entanglement of qumodes. Through this, it is possible to instead induce the desired non-Gaussian characteristics, paving the way for simulations that capture non-trivial quantum dynamics.

A central aspect of quantum systems that one might wish to explore using such methods is the Hamiltonian and the time-evolution that is governed by it, a cornerstone in understanding the dynamics of quantum particles and fields~\cite{Jordan:2011ci, Jordan:2012xnu, Jordan:2014tma, PhysRevA.92.063825, Jordan_2018, Klco:2018zqz,Banuls:2019bmf}. By simulating the time evolution governed by a system's Hamiltonian, we can explore how quantum states change over time, which is crucial for predicting the behaviour of quantum particles and systems under various conditions. This process is essential for simulating inherently quantum-mechanical phenomena that cannot be accurately modelled using classical physics. One of the most notable examples is quantum tunnelling, a phenomenon where particles pass through potential barriers that would be insurmountable, according to classical mechanics. This process is critical in a wide range of quantum systems, from the decay of atomic nuclei to the operation of quantum dots and superconducting qubits. By simulating the time evolution of quantum systems, one can also investigate other observables, such as energy spectra, correlation functions, and phase transitions, providing deep insights into the nature of quantum materials, chemical reactions, and even the evolution of early universe conditions in high-energy physics. CVQC offers a novel approach to simulating the time evolution of quantum states under arbitrary Hamiltonians. By decomposing the time evolution into discrete steps through Trotterization, we show how CVQC facilitates the simulation of complex quantum systems, including those governed by non-Gaussian potentials. 

In contrast, quantum gate computing operates within a digital framework, encoding quantum information in discrete qubits and manipulating it through a sequence of quantum gates. While this approach has paved the way for significant advancements in quantum computing, and has gained interest for uses in the simulation of quantum field theories~\cite{Jordan:2011ci, Jordan:2012xnu, Jordan:2014tma, PhysRevA.92.063825, Jordan_2018, Klco:2018zqz,Banuls:2019bmf, Abel:2020qzm,Abel:2020ebj,PRXQuantum.4.027001, Kane:2022ejm}, high-energy particle collisions~\cite{Blance:2020ktp, Bepari:2020xqi,Bepari:2021kwv,Gustafson:2022dsq}, and machine learning~\cite{Schuld2015, Carleo:2019ptp,Abel:2022lqr, Abel:2023erv, Rousselot:2023pcj},  it inherently approximates the continuous nature of quantum systems, potentially limiting its ability to capture the full spectrum of quantum dynamics.
Thus, CVQC, with its continuous-variable approach, offers a promising avenue for simulating quantum systems complementary to quantum gate computing approaches.

The simulation of non-trivial Hamiltonians on CVQC devices has been difficult due to the challenges in implementing non-Gaussian operations on photonic devices. Most current approaches propose circuits which involve non-Gaussian gate operations generated by non-linear optics to achieve Hamiltonian simulation on a continuous-variable device~\cite{PhysRevA.92.063825, PhysRevA.97.062311}. However, experimentally such operations are difficult to produce~\cite{Lloyd1999, RevModPhys.77.513, Adesso_2014}. This has led to attempts to generate non-Gaussian effects through the use of Gaussian operations and measurements~\cite{PhysRevA.105.012412}, in particular References~\cite{PhysRevA.100.012326,PhysRevA.100.052301} utilise a machine-learning routine to enhance the production of non-Gaussian states. In this paper, we build and improve on this approach and propose a quantum circuit for simulating the Trotterised time-evolution of a quantum-mechanical state under the influence of an arbitrary Hamiltonian using only Gaussian operations and measurements. The circuit for a single Trotter step can then be applied iteratively to achieve the time-evolution of the Hamiltonian. By using a {\it top-hat} resource function and generating the {\it evolver-state}, the non-Gaussian part of the Trotter evolution, using a measurement-based circuit, we show that the effect of the {\it noise-factor} from Reference~\cite{PhysRevA.100.012326} can be maximally suppressed without reducing the strength of the overall operation. We demonstrate the circuit's ability to simulate time-evolution for several examples of quantum-mechanical Hamiltonians and find that the circuit performs remarkably well compared to exact, classical simulations. Furthermore, we show that the approach can be extended, outlining how the continuous behaviour of the CVQC device can be harnessed to accurately simulate quantum field theories, maintaining the continuous nature of fields. 

In Section~\ref{sec:CVQC} we outline the necessary background on optical quantum computing required for the rest of the paper. Section~\ref{sec:trottProp} details the circuit architecture for the Trotterised time-evolution of a quantum-mechanical state under the influence of an arbitrary Hamiltonian. The circuit is then tested in Section~\ref{sec:results}, comparing the output of a continuous-variable quantum simulator against an exact, classical calculation. Finally, in Section~\ref{sec:qft}, we explain how this method can be extended to simulate quantum field theories on a CVQC device.


\section{Background: Continuous-Variable Quantum Computing}\label{sec:CVQC}

The framework we will consider is quantum computation via quantum optics, which offers an experimentally realisable approach to Continuous-Variable Quantum Computing (CVQC), in which the continuous-variable system is constructed from the quantised electromagnetic field~\cite{RevModPhys.77.513, Adesso_2014}. The use of quantum optics as a method of quantum computing benefits from the exceedingly low decoherence properties of photons, and therefore holds good potential for transmitting and maintaining quantum information throughout many gate operations~\cite{Kok2007, Ladd2010,Bromley_2020}.  In this realisation, the Hilbert space of each qumode is the infinite-dimensional photon-number degree of freedom~\cite{EISERT2003} which is the Fock basis, such that the total Hilbert space, $\mathscr{H}$, of an $N$ qumode system is

\begin{equation}
\mathscr{H} = \bigotimes^N_{i=0} \mathscr{H}_i~,
\end{equation}
where $\mathscr{H}_i$ is the Hilbert space of the \nth{i}  qumode. The system can then be modelled as a set of $N$, non-interacting, quantum harmonic oscillators. Each qumode is an oscillator based on the simple-harmonic-oscillator (SHO) Hamiltonian,
\begin{equation}\label{eqn:QHO}
\mathcal{H}_{\rm SHO} = \frac{1}{2} \left(\frac{\hat{p}^2}{m} + m\omega^2\hat{x}^2\right)~,
\end{equation}
where the continuous operators $\hat{x}$ and $\hat{p}$ are defined in terms of the creation and annihilation operators, $\hat{a}^\dagger$ and $\hat{a}$ respectively, 
\begin{align}\label{eqn:xandpop}
m \omega \hat{x} &~=~ \sqrt{\frac{m\hbar \omega}{2}} (\hat{a} + \hat{a}^\dagger)~, &i\hat{p}~ =~ \sqrt{\frac{m\hbar \omega}{2}} (\hat{a} - \hat{a}^\dagger)~,
\end{align}
and obey the standard commutation relation
\begin{equation}
[\hat{x}_i, \hat{p}_j] = i\hbar \delta_{ij}.
\end{equation}
It will be convenient to express the state on a qumode as an expansion in the Fock basis, such that

\begin{equation}\label{eqn:evolve_focked_gen}
\vert \psi \rangle ~=~  \sum^{\infty}_{n=0} A_n \vert n \rangle~,  
\end{equation}  
where $A_n$ is the coefficient of the \nth{n} Fock state, $\left\vert n\right\rangle$. For the rest of this paper, we will adopt natural units, $\hbar=m=\omega=1$. 

In the CVQC framework, computation is achieved by applying quantum gate operations on the qumodes.  The simplest to achieve are the so-called {\it Gaussian} gates which act on the qumodes as quadratic phase operators in the quadratures, that is in full generality they take the form 
\beq
e^{i \theta_{ij} \hat x_i \hat x_j + i \theta'_{ij} \hat x_i \hat p_j +  i \theta''_{ij} \hat p_i \hat p_j}~,
\eeq
where $i$ and $j$ label the qumodes on which they act, and $\theta_{ij},\theta'_{ij}$ and $\theta''_{ij}$ are constants. For universal computation it is essential also to be able to implement {\it non}-Gaussian gates (i.e. with $\hat x^3$ and higher appearing in the phase) \cite{Lloyd1999,PhysRevA.100.012326,PhysRevA.100.052301}. This is a delicate process because the fact that photons do not strongly interact with each other, while being excellent for maintaining coherence,  is accompanied by the negative implication that non-Gaussian effects are hard to achieve. Directly producing non-Gaussian gate operations requires the use of non-linear optical materials which induce non-Gaussian effects, such as the Kerr effect. However, currently the non-Gaussian effects induced by known non-linear optical materials are extremely weak~\cite{RevModPhys.77.513, Slussarenko2019,Kok2007, Ladd2010, PhysRevA.100.052301}, and the generation of arbitrary non-Gaussian operations is difficult. To circumvent these difficulties an alternative is the measurement-based approach (see References~\cite{PhysRevA.100.012326,PhysRevA.100.052301} and references therein), which we will be using here. Here non-Gaussianity is introduced by a process of post-selected measurement on entangled qumodes (in other words the accepted photons are filtered by the result of the measurement on an entangled ancilla qumode). 

This Section will present the necessary gate operations of both kinds that will be required for our paper. In Section~\ref{sec:GaussianOps}, we review the action of the various {\it Gaussian} gate operations, and then Section~\ref{sec:nonGaussianOps} outlines the construction of non-Gaussian operators. Although the methods presented here are general, this paper will utilise the Gaussian gate operations available on the \texttt{StrawberryFields} platform~\cite{Killoran_2019, Bromley_2020}.

\subsection{Review of required Gaussian gate operations}\label{sec:GaussianOps}
 The required Gaussian gate operations in the continuous-variable regime are expressed either in terms of the quadrature operators, $\hat{x}$ and $\hat{p}$, or the creation and annihilation operators, $\hat a^\dagger$ and $\hat a$, of the conventional  SHO defined in Equation~\ref{eqn:xandpop}. Here we will outline the gates by highlighting the action that is most relevant for this paper (for a full set of expressions the reader is referred to Reference~\cite{RevModPhys.77.513}): 
\begin{itemize} 

\item{\it Squeezing}: The action of a squeezing gate with parameter $z=r e^{i\phi}$ is 
\beq
 S^\dagger (z) \hat x  S(z) ~=~ e^{-r} \hat x~;\qquad 
 S^\dagger (z) \hat p  S(z) ~=~ e^{r} \hat p~. 
\eeq 
Note that somewhat counterintuitively $S(r)$ maps the wavefunction in the $x$-basis as 
\beq
\label{eq:squ}
( S\psi)(x) ~=~ e^{r/2} \psi(e^r x)
\eeq
 where we maintain normalisation with the prefactor. (In detail for this one case, in Dirac notation we have $ S\ket{x} = \ket{e^{-r} x} $ so that $ ( S\psi)(x)  = \sandwich{x}{ S} {\psi} =  \sandwich{e^r x}{S^\dagger S} {\psi} = \braket{e^r x}{\psi}$).   
  
  \item{\it Displacement}: A displacement gate with complex parameter $\alpha$ has the following action on the $\hat x$ and $\hat p$ operators:
\beq
 D^\dagger (\alpha) \,\hat x  \,
 D (\alpha) ~=~ \hat x +\sqrt{2}\, \Re (\alpha)~ ~;\qquad D^\dagger (\alpha) \, \hat p  \,
 D (\alpha) ~=~ \hat p +\sqrt{2}\, \Im (\alpha)~,
 \label{eq:disp_gate}
\eeq 
which maps the wavefunction and its Fourier transform  as
\beq
( D\psi)(x) ~=~  \psi(x-\sqrt{2} \,\Re(\alpha) )~~; \qquad ( D\tilde \psi)(p) ~=~  \tilde \psi(p+\sqrt{2} \,\Im(\alpha) )~~.
\eeq

\item{\it Rotation}: The action of a rotation gate with real parameter $\theta$ is given by 
\beq\label{eq:rot}
 R (\theta )  ~=~ 
 e^{i \theta \hat a ^\dagger \hat a} ~=~ e^{i \theta \left(  \frac{1}{2} \hat p^2+ \frac{1}{2}  \hat x^2 - \frac{1}{2}\right)} ~.
 \eeq 
 As the phase $e^{-i\theta/2}$ corresponding to the ground-state energy acts universally, it will usually be possible to ignore it.

\item{\it Controlled-X}: We will in addition to the above single qumode operators be using several Gaussian two-qumode operators. These induce displacements in $x$ or $p$ of qumode-$x$ which depend on the value of $y$ or $p_y$ measured on a second qumode-$y$ to which it is coupled, and vice-versa. In terms of operators the controlled-X gate (again taking $\hbar =1$) for two qumodes with variables $x,p_x,y$ and $p_y$ is 
\beq
 C_X (s; \hat y, \hat p_x )  ~=~ 
e^{- i s \hat y \hat p_x } ~,
\label{eq:controlX}
 \eeq
 which sends $\hat x \to \hat x + s \hat y $. Using the same steps as for the squeezing gate, the action on a product wavefunction is $C_X (s; \hat y, \hat p_x )  \psi(x )\psi'(y ) =  \psi(x-sy ) \psi'(y )$.

\item{\it Controlled-Z}: The second type of control gate that will be needed for this discussion is the controlled-Z gate. In terms of operators the controlled-Z gate  for two qumodes with variables $x,p_x,y,p_y$ is 
\beq
 C_Z (s; \hat y, \hat x )  ~=~ 
e^{- i s \hat y \hat x } ~.
 \eeq

\end{itemize}
The final element that is required for the discussion is the notion of {\it homodyne measurement}. This is a 
projection of the state onto particular $x$ or $p$ values or a linear mix. In the Gaussian system this is done by projecting onto squeezed states, with a variance of $\sigma = 2 \times 10^{-4}$~\cite{Killoran_2019, Bromley_2020}. 

\begin{figure}[t!]
\centering
	\begin{subfigure}{0.49\textwidth}
	\centering
	\includegraphics[width=\textwidth]{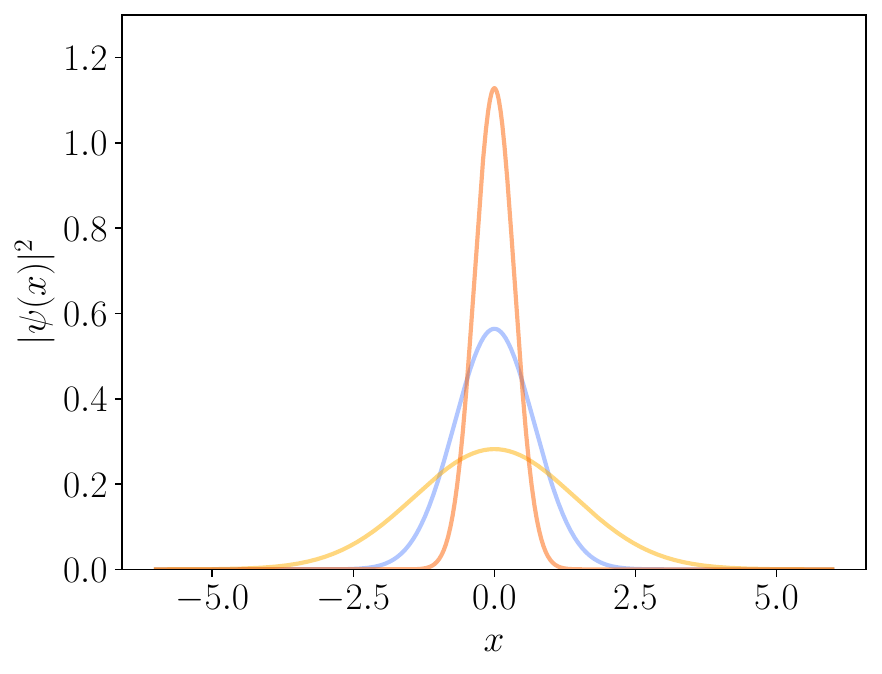}
	\caption{}\label{fig:testsa}
	\end{subfigure}
	\hfill
	\begin{subfigure}{0.49\textwidth}	
	\centering
	\includegraphics[width=\textwidth]{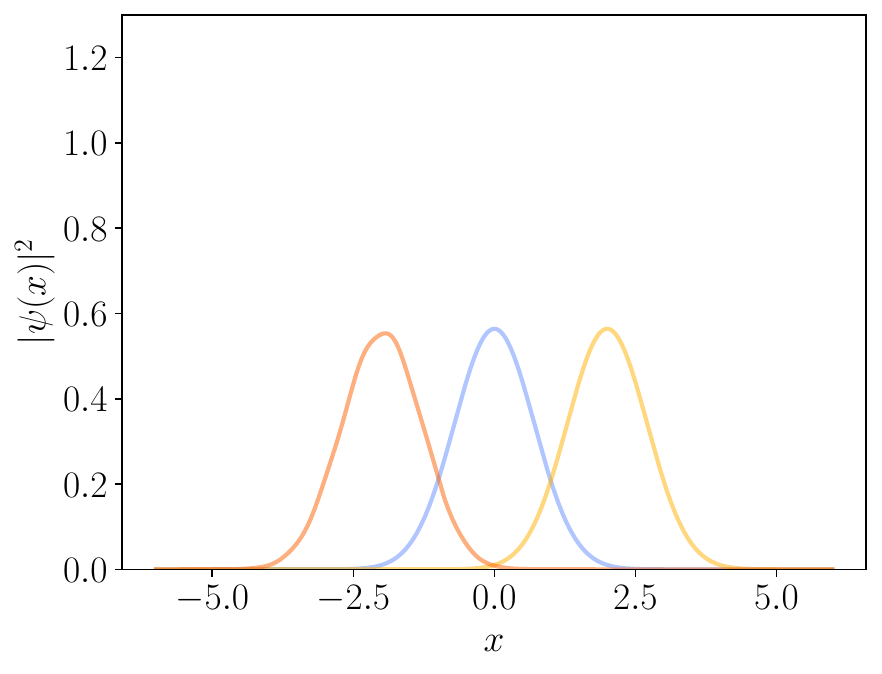}
	\caption{}\label{fig:testsb}
	\end{subfigure}
\caption{Example operations using a squeezing gate, and a controlled-X gate together with homodyne measurement. In (a) the ground state is squeezed by $S(\ln(1/2))$ which produces a wave-function flattened by a factor of 2 (yellow line). Then it is squeezed again by $S(\ln\,(4))$ producing a ground state squeezed by a factor 2 (orange line). In (b) we perform a composite displacement, by using a Controlled-X gate followed by a homodyne measurement of $y$, first displacing by 2 to the right, so that $\psi(x)\to \psi(x-2)$ (yellow line), then displacing by 4 to the left, so that $\psi(x-2)\to \psi(x+4)$ (red line). In this case the Fock truncation is 60, and we see some distortion beginning to appear at the peak.} \label{fig:tests}
\end{figure}

\begin{figure}[t!]
\centering
	\begin{subfigure}{0.49\textwidth}
	\centering
	\includegraphics[scale=0.7]{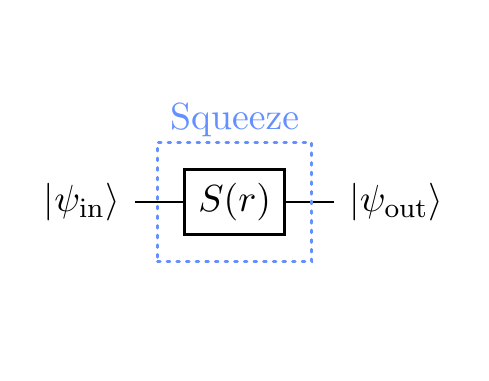}
	\caption{}\label{fig:simpcirca} 
	\end{subfigure}
	\hfill
	\begin{subfigure}{0.49\textwidth}	
	\centering
	\includegraphics[scale=0.7]{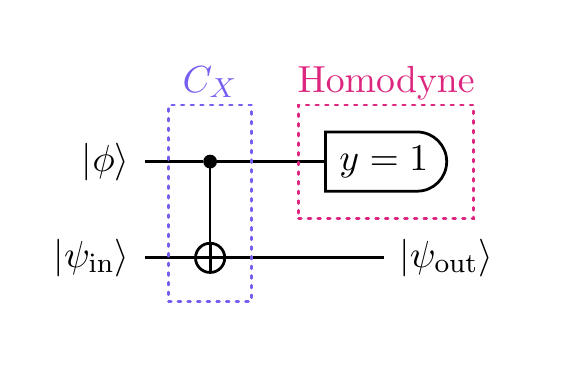}
	\caption{}\label{fig:simpcircb} 
	\end{subfigure}
\caption{Circuit diagram representation of the simple manipulations which produced Figure~\ref{fig:tests}. Following the convention for control gates in discrete gate systems, the controlled-X gate in (b) is controlled by the upper $y$ qumode (represented by a solid circle) and acts on the lower $x$ qumode (represented by cross-hairs).}\label{fig:simpcirc} 
\end{figure}

As a warm-up exercise, let us perform a few test manipulations utilising  some of these gates on a qumode groundstate. Figure~\ref{fig:testsa} shows the groundstate $\braket{x}{0}$ together with two manipulations corresponding to $S(\ln(1/2))\braket{x}{0}$ which produces a widened wave function (in yellow), and a 
second state corresponding to $$S(\ln(4))S(\ln(1/2))\braket{x}{0}$$ (in orange).  
The second panel, Figure~\ref{fig:testsb},  shows the ground state together with two displacement manipulations. These were performed not using the displacement gate in Equation~\eqref{eq:disp_gate}, but with the controlled-X gate of Equation~\eqref{eq:controlX} followed by a homodyne measurement on $y$. That  is we begin with two qumodes in their ground states, and act on it with a controlled-X gate, 
resulting in 
\beq 
\label{eq:CXact}
 C_X(s; \hat y, \hat p_x) \braket{x}{0}\braket{y}{0}~=~  \braket{x - s y}{0}\braket{y}{0} ~
 \eeq 
 and then in this example we perform a homodyne measurement at $y=1$ resulting in the displaced state $\braket{x - s}{0}$.
The first  case (yellow line)  takes $s=2$ resulting in $\psi(x)\to \psi(x-2)$ and the curve moves to the right. The second case (orange line) repeats the displacement operation with $s=-4$ resulting in the displacement $\psi(x-2)\to \psi(x+2)$ and the curve is then displaced to the left by 4 units. 
The two sets of manipulations which produced Figures~\ref{fig:testsa} and \ref{fig:testsb} were generated respectively by the two simple circuit diagrams shown in Figures~\ref{fig:simpcirca} and \ref{fig:simpcircb}.

\subsection{Creating non-Gaussian operations}\label{sec:nonGaussianOps}

As mentioned, achieving universal quantum computation with continuous-variable devices requires non-Gaussian operations generated by Hamiltonians of cubic order or higher in the $\hat{x}$ and $\hat{p}$ operators from Equation~\eqref{eqn:xandpop}~\cite{Lloyd1999, PhysRevA.100.012326, PhysRevA.100.052301}. Here the measurement-based approach of References~\cite{PhysRevA.100.012326, PhysRevA.100.052301} will be used. In this method Photon-Number-Resolving (PNR) measurements induce a non-Gaussian state on a target qumode. A disadvantage of moving to a measurement-based framework is that the production of the non-Gaussian operation is now probabilistic. However machine learning can be used to optimise the success of producing the {\it desired} non-Gaussian operation~\cite{PhysRevA.100.012326, PhysRevA.100.052301}. 

The circuit that will be trained to produce non-Gaussian states follows the $n$-qumode Gaussian Boson Sampling (GBS) architecture~\cite{Hamilton2017}, first transforming the qumode states to displaced-squeezed states by applying a series of displacement and squeezing operations to each of the qumodes. The system is then entangled by feeding the displaced-squeezed states through an interferometer, constructed using the rectangular arrangement of beamsplitters from Reference~\cite{Clements2016}. This process is then repeated for $I$ layers, with each layer being parameterised with trainable variables, $\theta_i$. The number of layers depends on the required expressibility of the circuit. Finally $(N-1)$-post-selected-measurements are made using PNR detectors to induce a non-Gaussian state on the target qumode, $\vert \phi \rangle$.

\begin{figure}
\centering
\begin{subfigure}{0.49\textwidth}
\centering
\includegraphics[width=\textwidth]{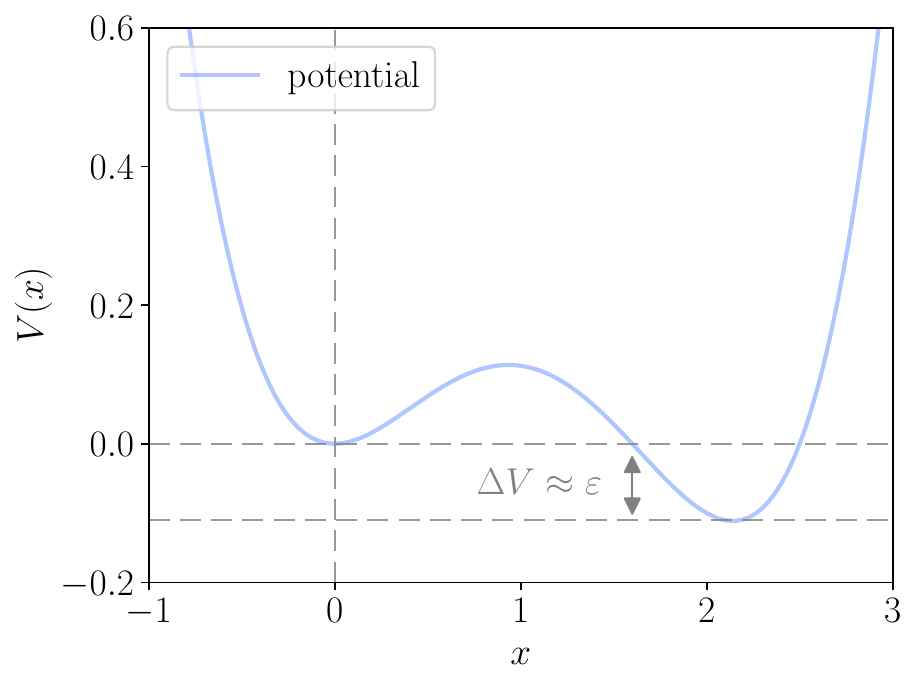}
\caption{}\label{fig:potentiala}
\end{subfigure}
\hfill
\begin{subfigure}{0.49\textwidth}
\centering
\includegraphics[width=\textwidth]{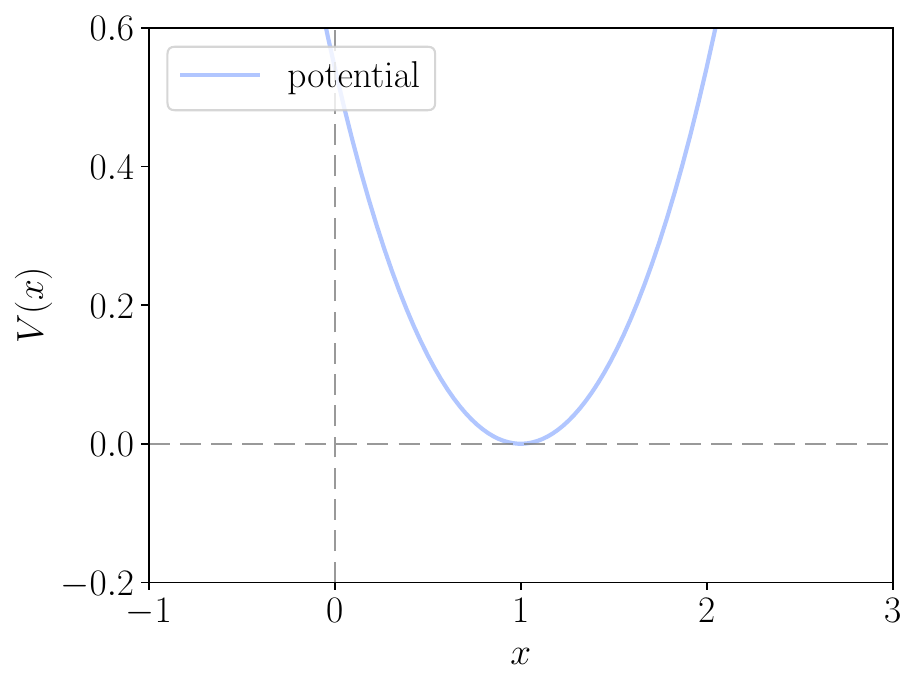}
\caption{}\label{fig:potentialb}
\end{subfigure}
\begin{subfigure}{0.49\textwidth}
\centering
\includegraphics[width=\textwidth]{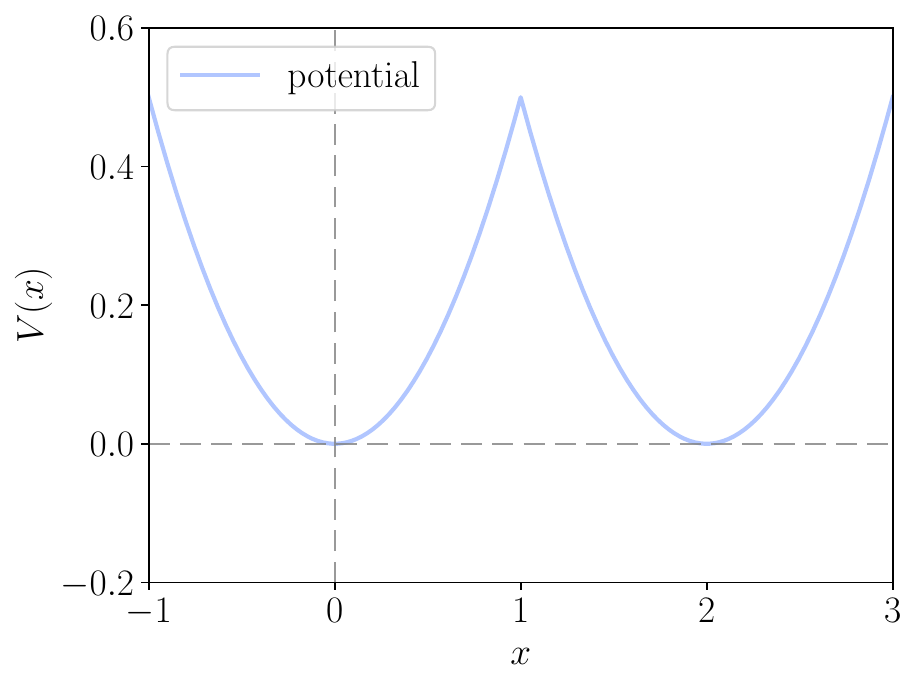}
\caption{}\label{fig:potentialc}
\end{subfigure}
\caption{The potentials of focus in this study: namely the (a) quartic, (b) $\cosh$ potential and (c) `$w$' potential. In all cases the wave-function is initialised as a Gaussian state centred on the origin.}\label{fig:potential}
\end{figure}

The circuit parameters are then trained by a classical machine learning routine to fit the output of the circuit to a target state vector. At each step of the training, the loss is calculated by truncating the Fock state expansion from Equation~\eqref{eqn:evolve_focked_gen}. The loss has the form 

\begin{equation}
\mathcal{L} = \frac{1}{n_{\rm max}} \sum_{n=0}^{n_{\rm max}} \left| A_n - A^\prime_n \right|  ^2~.
\end{equation}
where $A_n$ and $A_n^\prime$ are the \nth{n} coefficients of the trained state and target state respectively in the Fock basis, and $n_{\rm max}$ is the truncation.


\section{Implementing time-evolution of wavefunctions}
\label{sec:trottProp}

Having collected the required ingredients, we now present a Continuous-Variable Quantum Computing (CVQC) framework for simulating the time-evolution of quantum states under the influence of an arbitrary potential. Section~\ref{subsec:principle} describes the principle whereby an arbitrary state can be made to evolve under the influence of an arbitrary Hamiltonian, by using an ancilla qumode initialised with a prescribed non-Gaussian state called the {\it evolver}-state. Then Section~\ref{sec:ML} discusses how the evolver state may itself be constructed using the measurement-based approach described in  Section~\ref{sec:nonGaussianOps}. 

\subsection{Schr\"odinger evolving wavefunctions with arbitrary Hamiltonians }

\label{subsec:principle}

Let us start by explicitly stating the goal. In a non-relativistic system, the evolution of any quantum state  is driven by the Hamiltonian, which takes the following form (where recall that  $m=\hbar =1$ throughout): 
\begin{equation}\label{eqn:ham}
 {\mathcal{H}} ~=~ \frac{\hat p^2}{2} + V(\hat x)~.
\end{equation}
The goal is to be able to evolve any arbitrary input state  $\vert \psi_{\rm in}  \rangle$ under the influence of the Hamiltonian, ${\mathcal{H}}$, for which one must devise a photonic circuit that will implement the Schr\"odinger evolution
\begin{equation}\label{eqn:evolve}
\vert \psi _{\rm out} \rangle ~=~ e^{-i{\mathcal{H}}t} \vert \psi_{\rm in} \rangle ~. 
\end{equation}  
The techniques that will be developed here to do this are applicable to any potential, $V$, but in order to have a specific system in mind it is useful to focus on three specific cases. The first is the system with the potential 
\begin{align}
V(x) &~=~ \frac{1}{8} (x^2 - 2x)^2 - \frac{\varepsilon}{8} x^3, \nonumber\\
&~=~ \frac{1}{2} x^2 - \frac{(1 + \varepsilon/4)}{2}x^3 + \frac{1}{8} x^4~.
\label{eq:pot}
\end{align}
The depth of the true vacuum of this potential is $V_{\rm min}\approx \varepsilon$, at leading order. The potential is shown in Figure~\ref{fig:potentiala}. Around the origin this potential approximates the simple harmonic oscillator, thus the Gaussian SHO ground state at $x=0$ is an approximate energy eigenstate which will partially decay by tunnelling through the barrier.

\begin{figure}
\centering
\includegraphics[width=0.8\textwidth]{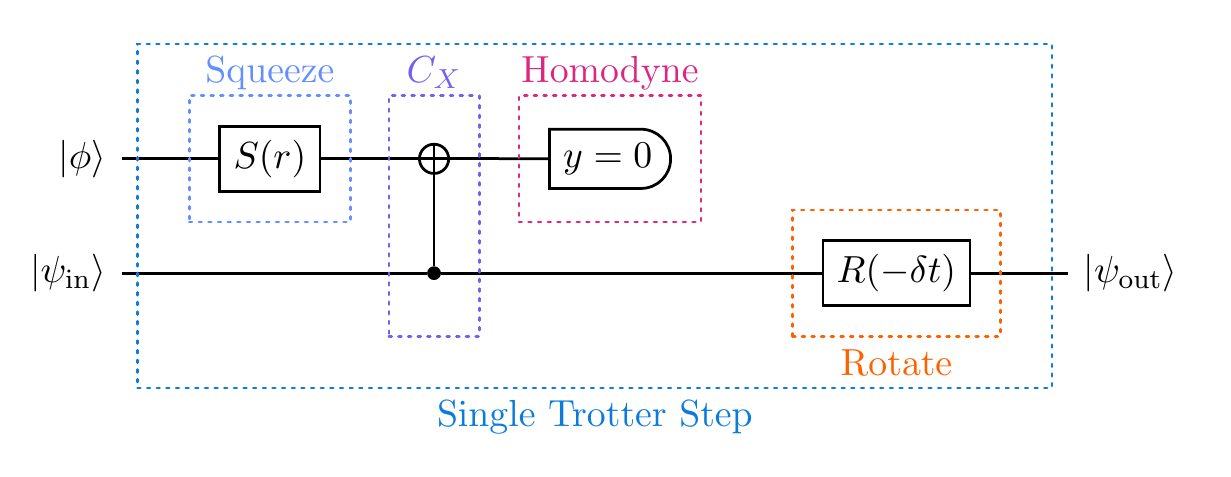}
\caption{Evolver-gadget to evolve through a single Trotter step. Here $\ket \phi $ is the evolver-state which is set according to Equation~\eqref{eq:castamps}, and which in Section~\ref{sec:ML} will be machine learned using a measurement-based quantum algorithm.}\label{fig:gadget}
\end{figure}

The second potential that will be considered is 
\beq 
V(x) ~=~ \cosh (x-1)-1 ~. 
\label{eq:cosh}
\eeq
This potential is of interest because its expansion around $x=1$ is   $V(x)=(x-1)^2/2+\ldots $, so to quadratic order it is also the SHO potential. Therefore any deviation from SHO behaviour is directly attributable to the higher-order terms, and moreover this deviation comes from a potential that is not polynomial. 
The final potential we will consider is 
\beq 
V(x) ~=~ \frac{1}{2}\left(\Theta(1-x) \times x^2 + \Theta(x-1) \times (x-2)^2 \right)~,
\label{eq:w_pot}
\eeq
where $\Theta(x)$ is the Heaviside function, which is shown in Figure~\ref{fig:potentialc}. This `$w$' potential is interesting for a number of reasons. First is the fact that the barrier in the potential is not differentiable, so there is not even a good polynomial approximation for it. 
The second reason this potential is interesting emerges upon considering solutions to the Schr\"odinger equation in a potential where the second minimum is missing and instead replaced with a flat region extending from the peak, {\it i.e.~} $V(x) ~=~ \frac{1}{2}\left(\Theta(1-x) \times x^2 + \Theta(x-1) \times 1 \right)$. In this potential the wavefunction is a bound state and none of it can escape to the right. In other words the ground-state energy of the SHO is smaller than the height of the barrier in the `$w$' potential. Therefore any barrier penetration in the full `$w$' potential of Eq.~\eqref{eq:w_pot} is entirely due to quantum tunnelling, and the behaviour that will be recovered is characteristically ``quantum''.

To implement the Schr\"odinger evolution of the wavefunction in Equation~\eqref{eqn:evolve}, there are two approximations that will be made. The first is to approximate the Fock expansion from Equation~\eqref{eqn:evolve_focked_gen}. That is
\begin{equation}\label{eqn:evolve_focked}
\vert \psi _{\rm out} \rangle ~=~ e^{-i\mathcal{H}t} \vert \psi_{\rm in} \rangle ~\approx~ \sum^{n_\textrm{max}}_{n=0} A_n(t) \vert n \rangle~,  
\end{equation}  
where $A_n$ is the  coefficient of the Fock state $\vert n \rangle$, and $n_\textrm{max}$ is the Fock state truncation. Thus in principle it is possible to determine the evolution in terms of a matrix acting on the Fock state coefficients, $A_n$, by expanding $\hat x$ and $\hat p$ in terms of the creation and annihilation operators as in Equation~\eqref{eqn:xandpop}.
Although such Fock truncation is not strictly speaking required for a particular photonic circuit to work, it {\it is} required to determine the parameters of the circuit itself, as will become clear. 

The second approximation that will be made is to Trotterise the time evolution, in other words to divide the total evolution time $t$ into $N$ steps of time $\delta t = t/N$. To do this it is convenient to separate out the Gaussian $\hat p^2/2+\hat x^2/2 $ part of the Hamiltonian because its contribution to the evolution can easily be generated by the rotation gate in Equation~\eqref{eq:rot}. That is, letting  
\begin{equation}
{ {\cal H}}~=~ H_0(\hat p,\hat x) + H_{1}(\hat x)~,
\end{equation}
where $H_0 = \frac{1}{2} (\hat p^2+\hat  x^2)$ is the Hamiltonian of the simple harmonic oscillator, and where
\begin{align}
H_1(x) &~=~  V( x)-\frac{ x^2}{2} \nonumber \\
&~=~ - \frac{(1 + \epsilon/4)}{2}x^3 + \frac{1}{8} x^4~,
\end{align}
 is the non-Gaussian part of the potential, 
  the operator $e^{-i  {\cal H} \delta t }$ corresponds to 
 \beq
 \label{eq:trotter_step}
 e^{-i  {\cal H} \delta t }~=~ R(-\delta t) ~ e^{-i H_1 (\hat x)\delta t +i {\cal O} ( \delta t^2)} ~.
  \eeq
 The complete evolution may then be enacted by applying $N$ of these so-called Trotter steps, 
 \be
\ket{\psi_{\rm out}}~=~   \left[ R(-\delta t) ~ e^{-i H_1 (\hat x)\delta t }\right]^N  \ket{\psi_{\rm in}}~.
  \ee
The $\delta t^2$ error in the Trotterisation approximation alluded to in Equation~\eqref{eq:trotter_step} arises because $H_0$ and $H_1$ do not commute, and it is given by the  Zassenhaus relation, 
 $$
 e^{A+B} ~=~e^Ae^Be^{
 [B,A]/2 +\ldots } ~,$$
where the dots denote higher-order commutators. Assuming that $[ \hat p^2 , H_1(\hat x)]\sim 1$, for such a trotterised evolution one therefore finds that an error accumulates  in the exponent of order $N \delta t^2$ and hence one requires $  \delta t \ll 1/t$ in natural units for the evolution to be accurate. Note that the Trotterisation error is retained  as a product of unitary operators, so that it is also unitary\footnote{It could in principle be improved by subtracting the leading $\delta t^2$ error with more compound Trotter steps, however this will not be necessary.}.
 
\begin{figure}
\centering
\includegraphics[width=\textwidth]{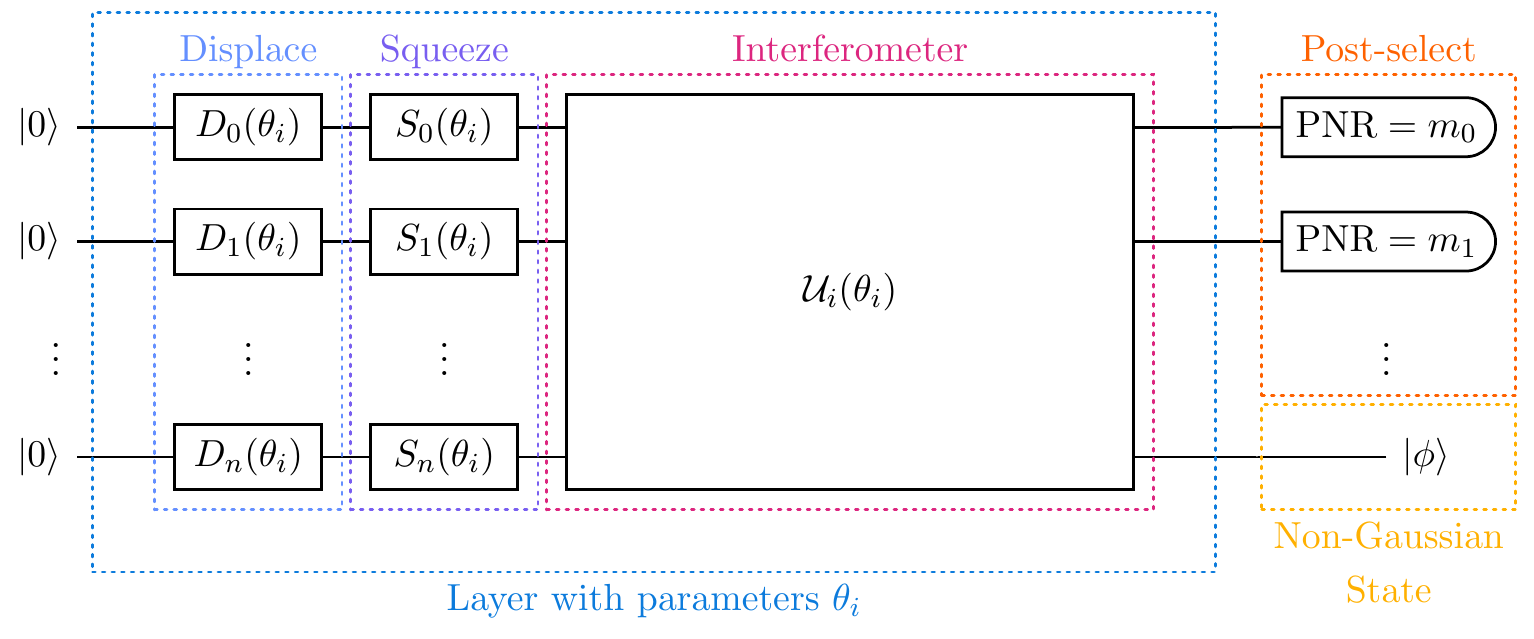}
\caption{Schematic of a quantum circuit for the preparation of a non-Gaussian state. The circuit architecture is inspired by a Gaussian Boson Sampling routine on $n$-qumodes. The incoming vacuum states are displaced, then squeezed before being interfaced with an interferometer, constructed using the rectangular architecture from Reference~\cite{Clements2016}. This routine is repeated for $I$ layers, parameterised with trainable variables, $\theta_i$, for each layer. Finally, ($n-1$)-measurements are made using Photon Number Resolving (PNR) detectors, with the $j$th measurement post-selecting on $m_j$. These measurements generate the induced non-Gaussian state on the \nth{n} qumode.}\label{fig:NGG}
\end{figure}

Thus the main task is to prepare a non-Gaussian operator that can act on an arbitrary state to give the  $e^{-i H_1 (\hat x)\delta t}$ factor in the Trotter step. For compactness of notation, consider the task of implementing an arbitrary non-Gaussian operation,
\beq
\label{eq:fphi}
  f(\hat x ) \, \ket{\psi_{\rm in}}~,
\eeq
as a circuit, where in this case  $f(\hat x)$ is  the desired unitary operation on $\ket{\psi_{\rm in}}$, 
\beq 
f(\hat x) ~\equiv ~ e^{-i H_1 (\hat x) \delta t  } ~.  
\eeq
Such a non-Gaussian operator can be constructed by improving the method presented in Reference~\cite{PhysRevA.100.012326}. The starting point of the method is to prepare  a state $\ket{\phi} $ on an ancilla qumode, with coordinate denoted $y$, which mirrors the desired Trotter step. In full generality an arbitrary state on the ancilla qumode can (since $f$ is invertible) be written as follows:
\beq
\label{eq:yphi0}
\braket{y}{\phi} ~=~ \sandwich{y}{f( \hat y/q )}{\phi_0}~, 
\eeq
where $\braket{y}{\phi_0}$ is some other resource state (to be determined) and where $q$ is a parameter whose role will become clear. 
This prepared $\braket {y}{\phi}$ state is the {\it non-Gaussian evolver-state}.

 The process of transferring the evolution to the input-state $\ket {\psi_{\rm in}}$ begins by entangling it with the evolver-state  using a controlled-X gate $\hat C_X(-s; x,y) \equiv  e^{ i s \hat x\hat p_y /\hbar }$ which induces a shift $ y \to  y + s x$ in the coordinates of the evolver-function as in Equation~\eqref{eq:controlX}, and implementing a squeezing  $
S(r;\hat y) $ with parameter $r$ chosen such that 
\begin{equation}
\label{eq:rsq}
e^r s ~=~ q~.
\end{equation} 
Next we make a rotation $R(-\delta t)$ on the $\ket \psi$ state and then finally the evolver-state is collapsed by making a homodyne measurement of $y=0$. The entire procedure is shown in the circuit diagram of Figure~\ref{fig:gadget}.

Consider the effect of this sequence of operations. Denoting the incoming state $\ket{\psi_{\rm in}}$  combined with the evolver-state $\ket{\phi}$  by a single ket, $\ket{\Psi}$,
 the output on the two qumodes after this sequence of operations, and before any measurements are made, can be written 
\beq 
\langle x,y|\Psi \rangle ~=~\bra{y}\bra{x} 
R(-\delta t;\hat x) \,  C_X(-s;\hat x,\hat y) \, S(r;\hat y)  \, f( \hat y / q) \,  |\phi_0\rangle \ket{\psi_{\rm in}}   ~.\eeq
According to Equations~\eqref{eq:squ} and \eqref{eq:CXact}, performing the various manipulations corresponding to these gates  and then performing the homodyne measurement $y=0$ with the choice of parameters in Equation~\eqref{eq:rsq}  yields an evolved state on the $\left\vert \psi\right\rangle$ qumode of the form  
\beq 
\langle  x|\psi\mbox{\small $(\delta t)$} \rangle ~=~
 \exp\left(-\frac{i}{2} (\hat p^2+\hat x^2 )\delta t \right) e^{ -i H_1 (  \hat x )\,\delta t } ~
   \langle  q   x |\phi_0\rangle \braket{x}{\psi_{\rm in}}   ~.
   \label{eq:trotter_sing}
   \eeq
This is the desired non-Gaussian evolution, corresponding to a single Trotter step, up to errors of order $\delta t^2$ and the {\it noise-factor} $ \langle  e^r s   x |\phi_0\rangle \equiv \braket{ x q }{\phi_0}$. 

All that remains is to choose the optimal form of the resource function $\braket{y}{\phi_0}$ in order to maximally suppress the effect of the noise function. There are two possibilities: one can choose a small value of $q$ which `freezes' the value of the function $\braket{x q}{\phi_0}$, and/or choose a flat resource function. It is convenient to adopt the top-hat function as the idealised resource function:
\be
\braket{y}{\phi_0}~=~ \begin{cases}  \frac{1}{L} & |y| <  L/2 \nonumber \\ 0 & |y| >  L/2~.\end{cases}
\ee
With this resource function the output state becomes 
\beq 
\ket{\psi \mbox{\small $(\delta t)$}} ~=~ 
\frac{ \Theta(L/2- q|\hat x| )}{L}~
 e^{ -i {\cal H} (  \hat x )\,\delta t }  \ket{\psi_{\rm in}} ~.
\eeq
Note that smaller values of $q$ allow larger domains in $x$ of valid evolution. 

Adopting this top-hat function as the resource state greatly simplifies the procedure, because one can determine all the Fock amplitudes of the evolver-state $\braket{y}{\phi} = \sandwich{y}{f( \hat y/q )}{\phi_0} $ directly, by numerically integrating  it against Fock modes:
\begin{align}
A^{(\rm evolver)}_n~\equiv~ \braket {n }{\phi} ~&
=~  \int_{-\infty}^{\infty}  \braket{ n }{y}  f( y/q) \braket{y }{\phi_0}dy ~,\nonumber \\
&=~ \frac{1}{L} \int_{-L/2}^{L/2}  e^{-i H_1(y/q)\delta t}  \braket{ n }{y} dy ~,\label{eqn:evolver}
\end{align}
where $\braket {y}{n}$ is the \nth{n}  Fock mode. Hence the evolver-state with which  the ancilla qumode must be initialised is 
\begin{align}
 \braket{ y }{\phi} ~=~ \sum_{n=0}^{n_{\rm max} }    A^{(\rm evolver)} _n \braket {y }{n } .
 \label{eq:castamps}
\end{align}
Due to the truncation $n_{\rm max}$, this is of course an approximation to  the idealised $f(y/q)$ function, which is expected to improve with higher $n_{\rm max}$.

Provided that one is able successfully to initialise this evolver state, the Trotter step circuit of Figure~\ref{fig:gadget} can be repeated $N$ times to evolve the state through time $t$. Therefore  being able to set this initial form of the evolver-state is the final ingredient that is crucial to be able to implement the procedure on genuine photonic devices. We now turn to this aspect.

\subsection{Learning the evolver-state}\label{sec:hamQML}

\label{sec:ML}

\begin{figure}[t!]
\centering
	\includegraphics[width=\textwidth]{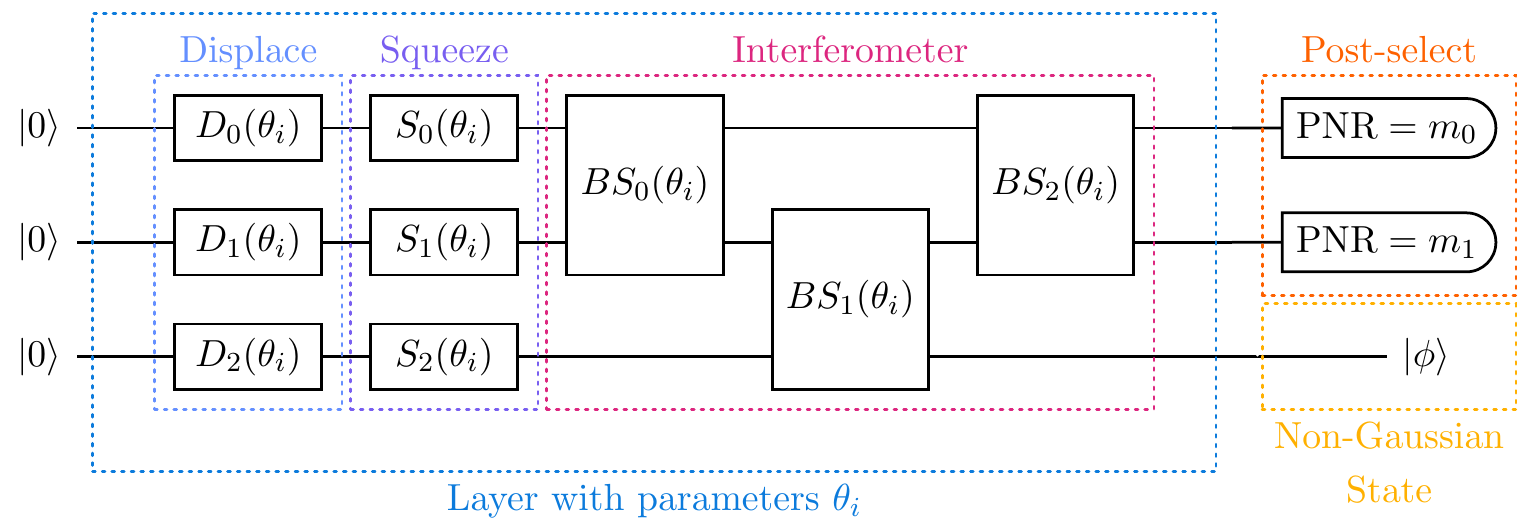}
\caption{Schematic of a three-qumode quantum circuit for the preparation of the non-Gaussian evolver-state.}\label{fig:3NGG}
\end{figure}

To fully implement the procedure outlined in Section~\ref{subsec:principle} on a photonic device, the non-Gaussian evolver-state, $\left|\phi\right\rangle$, must be prepared. Here, the measurement-based approach from Section~\ref{sec:nonGaussianOps} will be used, in which the circuit parameters of a circuit such as that in Figure~\ref{fig:NGG} are trained against the target state from Equation~\eqref{eq:castamps}. To achieve a good fit to the target, a circuit constructed from three-qumodes and 10 iterations of the layer method will be used. At each layer, the displaced-squeezed states pass through an interferometer which entangles the system.  Following in the rectangular architecture from Reference~\cite{Clements2016} only three beamsplitters are required in each layer for the three-qumode case.
Each gate operation has two parameters, thus the full circuit has 180 trainable variables. The training of the gate parameters has been restricted to values which are experimentally realisable~\cite{PhysRevLett.117.110801}. The values that the PNR measurements are post-selected on have not been included as  trainable parameters and have instead been chosen to be $m_0=m_1=5$. Reference~\cite{PhysRevA.100.052301} makes a detailed investigation into maximising success when creating non-Gaussian states using measurement-based quantum computing approaches. A schematic of the circuit diagram used to create the desired resource state for the time-evolution of an arbitrary Hamiltonian is shown in Figure~\ref{fig:3NGG}.

\begin{figure}[t!]
\centering
\begin{subfigure}{0.49\textwidth}
\includegraphics[width=\textwidth]{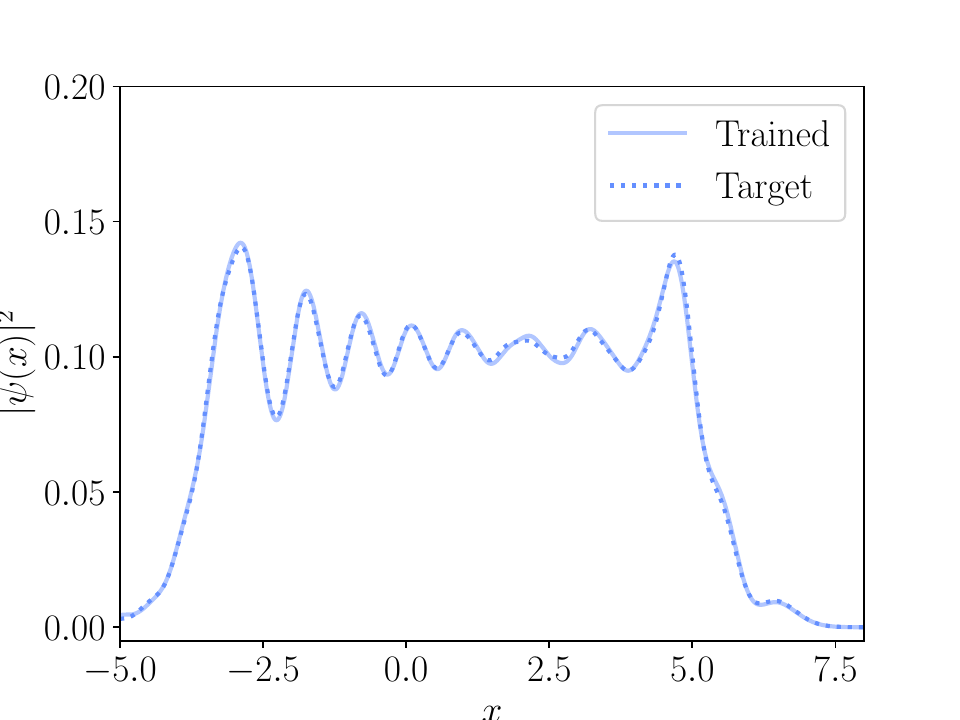}
\caption{}\label{fig:diea}
\end{subfigure}
\hfill
\begin{subfigure}{0.49\textwidth}
\includegraphics[width=\textwidth]{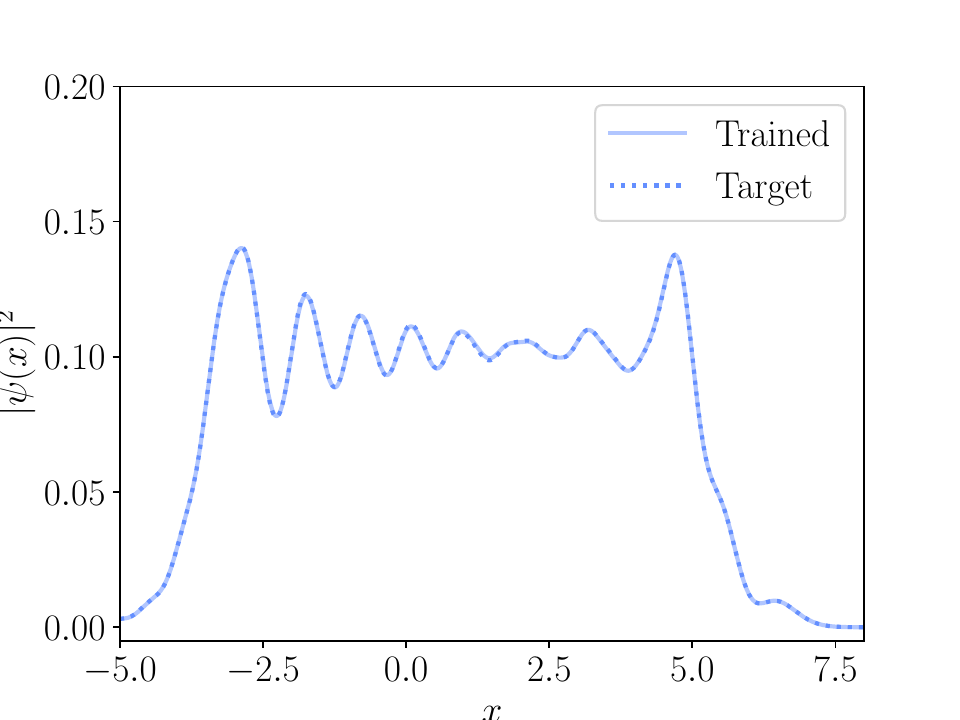}
\caption{}\label{fig:dieb}
\end{subfigure}
\caption{Normalised evolver-state function with a truncation to the first 25 Fock levels. The state has been trained using an $N$-qumode Gaussian Boson Sampling architecture with $(N-1)$-measurements, schematically shown in Figure~\ref{fig:NGG}, to produce the evolver-state on the \nth{N} qumode. }\label{fig:die}
\end{figure}

Figure~\ref{fig:diea} shows the state produced by the trained quantum circuit against the target from Equation~\eqref{eq:castamps} for the potential from Equation~\eqref{eq:pot} up to a Fock truncation of $n_{\rm max}=25$. The circuit achieves a good fit to the target state, however some discrepancies are visible in the region $1<x<3$. It is possible to increase the circuit size to four-qumodes, thus increasing the number of trainable parameters in 10 layers to 280. Figure~\ref{fig:dieb} shows the fit from such a  four-qumode circuit. One can see that the discrepancies are no longer visible and the fit to the target state is virtually exact. 

Although the training of the circuit can be a lengthy process, for each system the circuit only ever needs to be trained once to determine the circuit parameters for a given potential and a given $\delta t$, because the evolver-state is the same at each Trotter step. Once these parameters have been determined, the trained circuit can then be incorporated into the circuit to  generate the time-evolution of the wavefunction from Section~\ref{subsec:principle}.  In Section~\ref{sec:results}, the time-evolution simulated by the quantum circuit will be compared to an exact, classical calculation. 

\section{Results: Quantum mechanics on photonics versus numerically evaluated quantum mechanics}\label{sec:results}

In Section~\ref{sec:trottProp}, the quantum algorithm for the simulation of the Trotterised time-evolution of a wavefunction under a Hamiltonian with an arbitrary potential was proposed for a Continuous-Variable Quantum Computing (CVQC) approach utilising currently achievable quantum optics. The system builds a non-Gaussian evolver-state on an ancillary qumode using the measurement-based circuit from Section~\ref{sec:nonGaussianOps}, the parameters of which have been trained using a classical machine learning technique. In this Section, it will be shown that the algorithm performs as expected by investigating the evolution of a quantum-mechanical wave-function under the influence of the potentials from Equations~\eqref{eq:pot} and \eqref{eq:cosh}. Due to the excessive memory required to simulate circuits with more than four qumodes at a high Fock truncation, part of this study will be performed using the {\tt Ket} command from {\tt StrawberryFields}~\cite{Killoran_2019, Bromley_2020} to simply set the evolver-state, without the need of addition qumodes. 

\begin{figure}[t!]
\centering
\begin{subfigure}{0.49\textwidth}
\centering
\includegraphics[width=\textwidth]{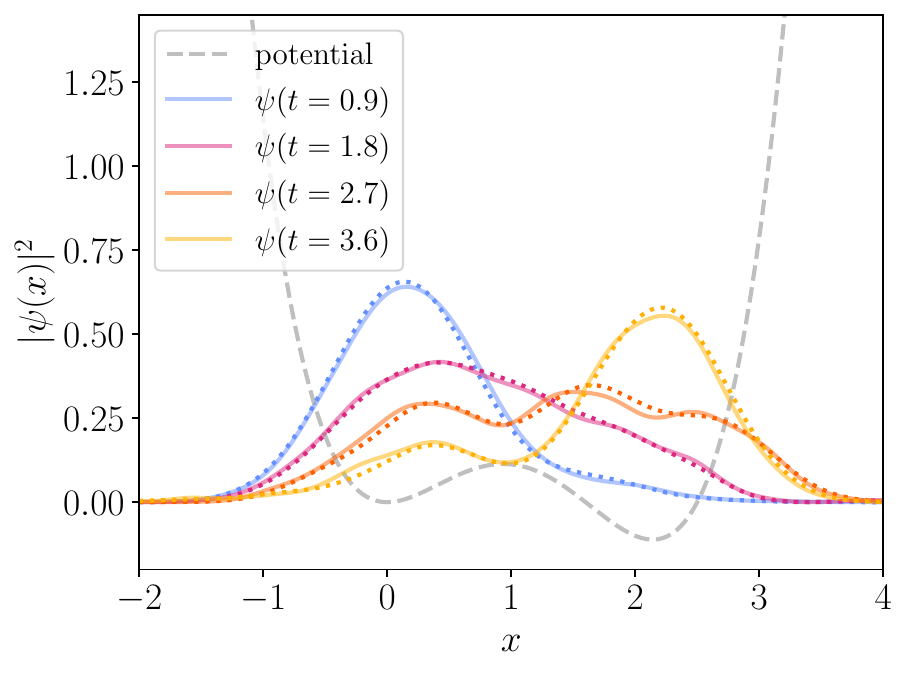}
\caption{}
\end{subfigure}
\hfill
\begin{subfigure}{0.49\textwidth}
\centering
\includegraphics[width=\textwidth]{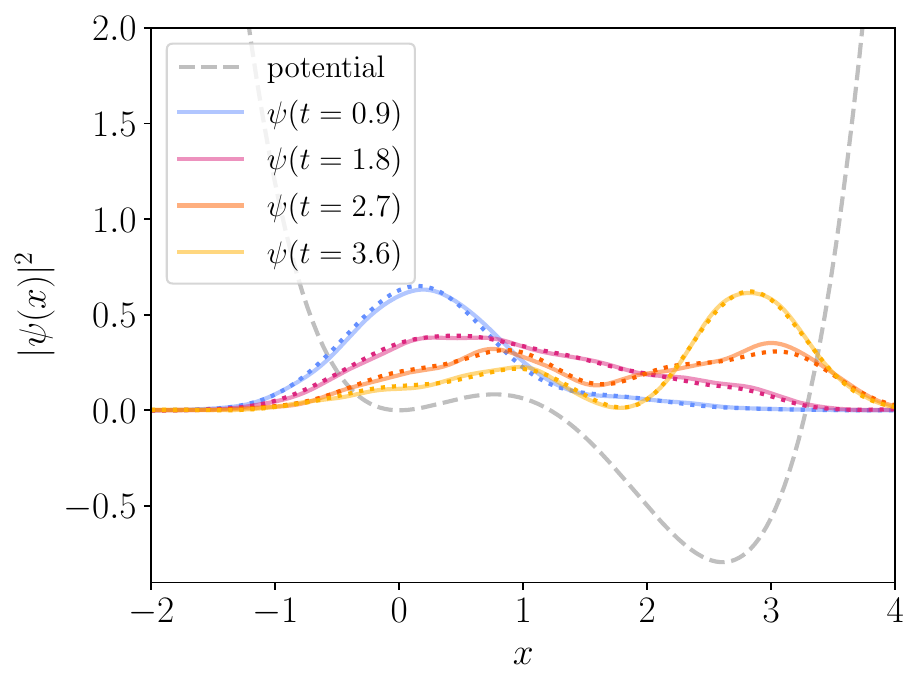}
\caption{}
\end{subfigure}
\caption{The time-evolution of a quantum system with the asymmetric quartic-potential of Equation~\eqref{eq:pot} with (a) $\varepsilon =0.1$ and (b) $\varepsilon =0.5$, generated by the photonic quantum simulator with a Fock truncation of 60 (solid line) and compared to an exact calculation (dotted lines).}\label{fig:Evolution}
\end{figure}

\begin{figure}[t!]
\centering
\begin{subfigure}{0.49\textwidth}
\centering
\includegraphics[width=\textwidth]{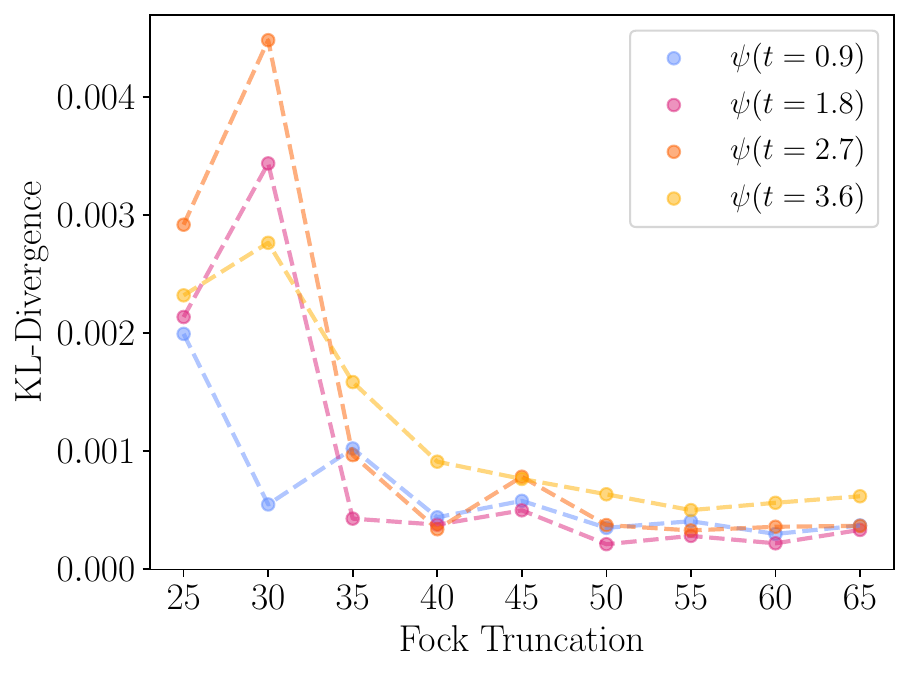}
\caption{}\label{fig:KLDiva}
\end{subfigure}
\hfill 
\begin{subfigure}{0.49\textwidth}
\centering
\includegraphics[width=\textwidth]{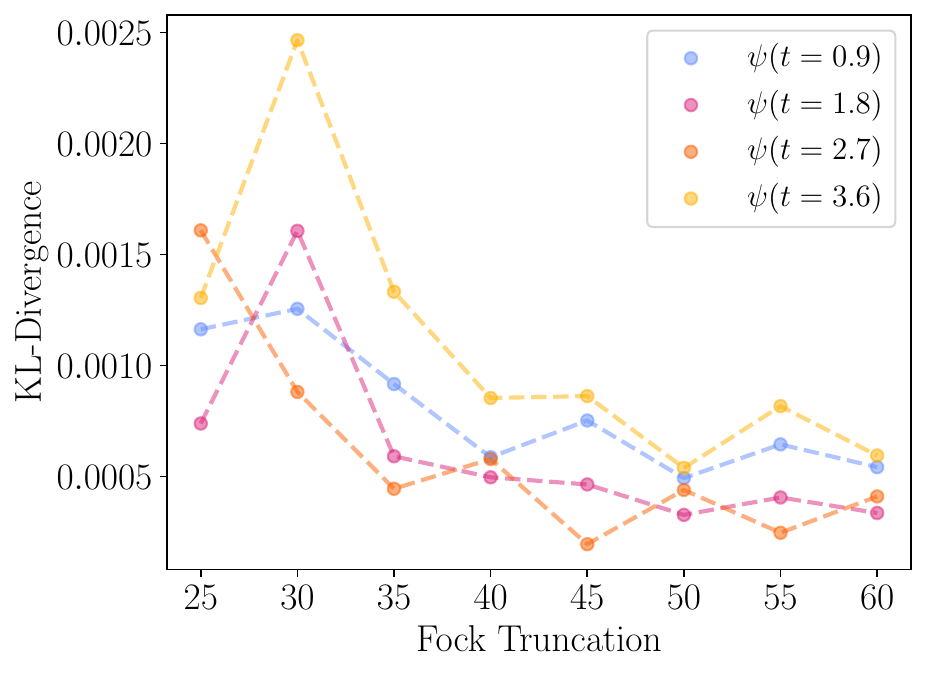}
\caption{}\label{fig:KLDivb}
\end{subfigure}
\begin{subfigure}{0.49\textwidth}
\centering
\includegraphics[width=\textwidth]{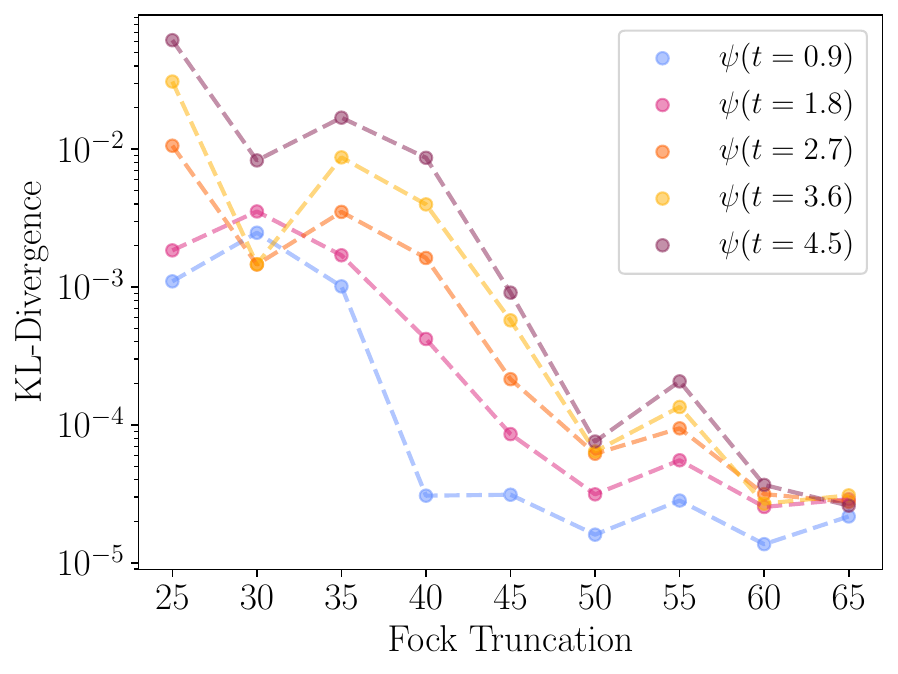}
\caption{}\label{fig:KLDivc}
\end{subfigure}
\caption{The Kullback-Leibler (KL) divergence between the quantum simulation and exact calculation for different evolution times and Fock truncations for the (a) quartic, (b) $\cosh$ and (c) `$w$' potentials. The KL divergence quantifies the disparity between probability distributions as a relative entropy (which broadly speaking encodes the information required to get from one distribution to the other). After a sufficient cutoff, the KL divergence exhibits a monotonic behaviour with time.}\label{fig:KLDiv}
\end{figure}

First, consider the potential from Equation~\eqref{eq:pot}, as shown in Figure~\ref{fig:potentiala}. The system is initialised in the ground state of the SHO, i.e. a Gaussian wavefunction centred around $x~=~0$. Figure~\ref{fig:Evolution} shows the time-evolution of the quantum-mechanical wavefunction simulated by the CVQC circuit (solid lines) compared to a classical simulation produced using {\tt Qibo}~\cite{qibo_paper} (dotted lines) for two scenarios: $\varepsilon =0.1$ and $\varepsilon =0.5$. The simulations have been run with a Trotter time-step of $\delta t=0.1$. Here, the {\tt Ket} command has been used to simulate the time-evolution at a Fock truncation of $n_{\rm max}=60$ on the quantum device. The agreement of the quantum algorithm with the classical simulation has been quantified using the Kullback-Leiber (KL) divergence~\cite{kullback1951information}, and is shown in Figure~\ref{fig:KLDiva} for the $\varepsilon=0.1$ case. It can be seen that, above a Fock truncation of $n_{\rm max}=35$, the agreement between the quantum and classical cases is remarkably good, and degrades monotonically with time  as one would expect given the accumulating Trotter and noise-factor errors discussed around Equation~\eqref{eq:trotter_sing}.

\begin{figure}[t!]
\centering
\begin{subfigure}{0.49\textwidth}
\centering
\includegraphics[width=\textwidth]{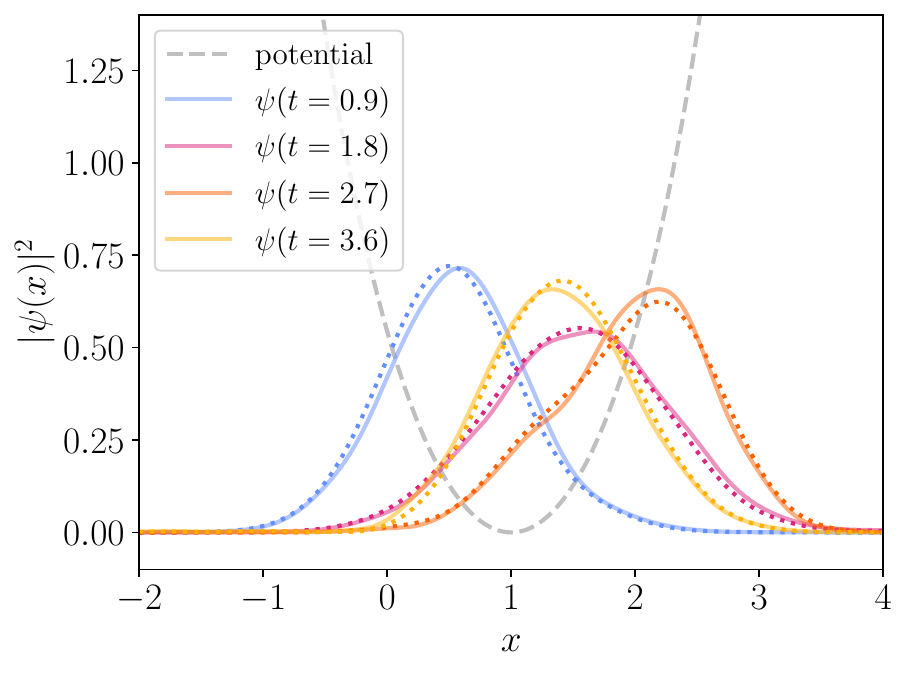}
\caption{}\label{fig:coshEvolvea}
\end{subfigure}
\hfill
\begin{subfigure}{0.49\textwidth}
\centering
\includegraphics[width=\textwidth]{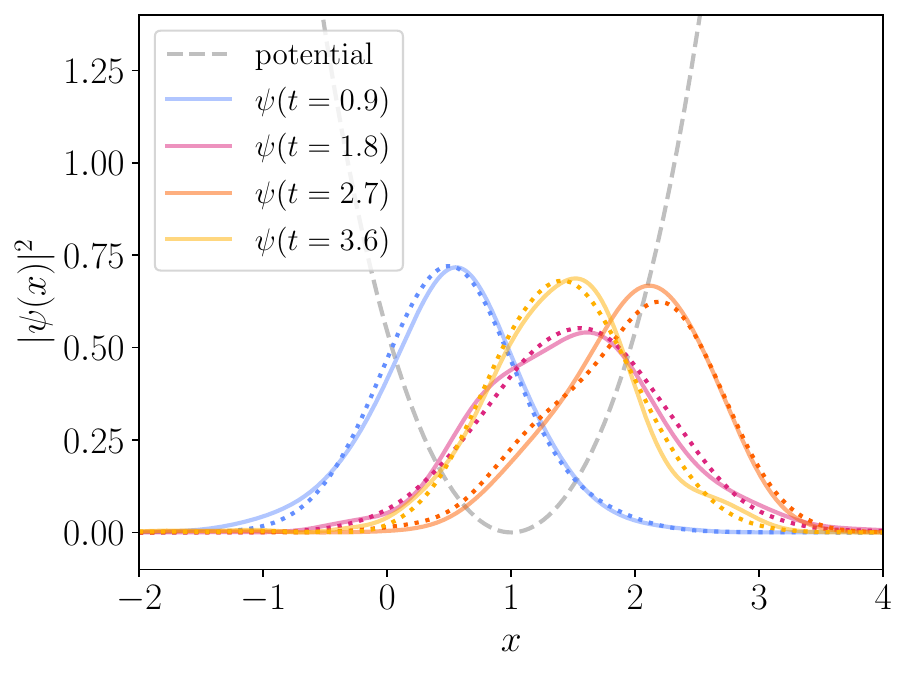}
\caption{}\label{fig:coshEvolveb}
\end{subfigure}
\caption{The time-evolution of a quantum system under the influence of the hyperbolic potential from Equation~\eqref{eq:cosh} generated by the photonic quantum simulator (solid lines) compared to an exact calculation (dotted lines). In (a) the circuit has used the {\tt Ket} command to initialise the evolver-state and has been run at a Fock truncation of 60. In (b) the evolver-state has been initialised using the full circuit and has been run at a truncation of 25.}\label{fig:coshEvolve}
\end{figure}

The second example that we consider is the time-evolution under the hyperbolic potential from Equation~\eqref{eq:cosh} shown in Figure~\ref{fig:potentialb}. Once again, the system is initialised in the ground state of the SHO Hamiltonian centred at the origin. Figure~\ref{fig:coshEvolvea} shows the comparison between the time-evolution of the wavefunction simulated by the quantum device and the classical device, with a Trotter time-step of $\delta t =0.1$ and a Fock truncation of $n_{\rm max}=60$ on the quantum simulation. To achieve the simulation up to a truncation of 60 the {\tt Ket} command has been used. The quantum circuit performs well, with the KL divergences showing good agreement for truncations greater than 35, as shown in Figure~\ref{fig:KLDivb}.

\begin{figure}[t!]
\centering
\includegraphics[width=0.76\textwidth]{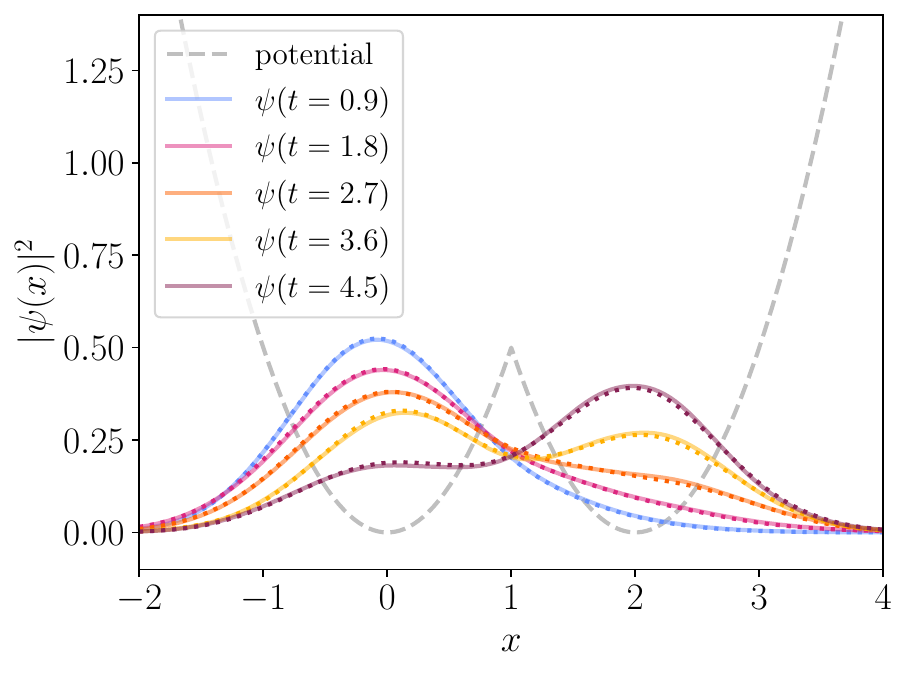}
\hfill
\caption{The time-evolution of a quantum system under the influence of the `$w$' potential of Equation~\eqref{eq:w_pot} generated by the photonic quantum simulator (solid lines) compared to an exact calculation (dotted lines).}\label{fig:wEvolve}
\end{figure}

Finally we consider evolution in the `$w$' potential of Eq.~\eqref{eq:w_pot} shown in Figure~\ref{fig:potentialc}. Again, the system is initialised in the ground state of the SHO Hamiltonian centred at the origin. Figure~\ref{fig:wEvolve} shows the comparison between the time-evolution of the wavefunction simulated by the quantum device and the classical device, with a Trotter time-step of $\delta t =0.1$ and a Fock truncation of $n_{\rm max}=60$ on the quantum simulation. 
The evolution is extraordinarily accurate with this potential. Indeed the KL divergences, which are shown in Figure~\ref{fig:KLDivc}, are extremely small for a sufficiently large Fock truncation. 

It is interesting to ask why the evolution in the `$w$' potential should be so much more accurate. Recall that the evolution in this case is expected to initially be dominated by tunnelling, implying that the penetration of the barrier (of height $V(1)=0.5$) is driven by exponential tails of the wavefunction. This in turn implies that the whole wavefunction is a bound state of the double well that must be exponentially suppressed beyond $x<-1$ and $x>3$. It is therefore insensitive to the edges of the top-hat resource state (which are well outside this range) and the noise-factor $\langle qx|\phi_0\rangle $ appearing in Eq.~\eqref{eq:trotter_sing}, in contrast with the situation in the other two potentials.

\begin{figure}[t!!]
\centering
\includegraphics[width=0.76\textwidth]{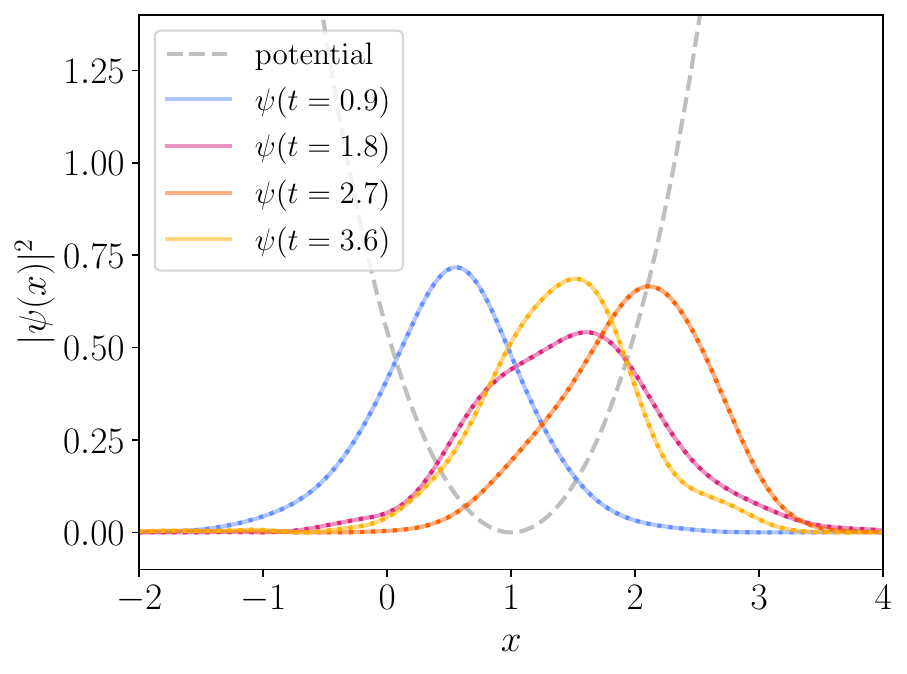}
\caption{A comparison between the time-evolution simulated using the full circuit (solid lines) and the {\tt Ket} command (dotted lines) for a Fock truncation of $n_{\rm max}=25$.}\label{fig:fullvsket}
\end{figure}

To fully test the performance of the proposed quantum algorithm, the {\it full circuit} was constructed for the cosh potential of Eq.~\eqref{eq:cosh}. Due to the memory constraints on the simulation, this circuit was  run using the three-qumode evolver-state preparation circuit from Figure~\ref{fig:3NGG} with a Fock truncation of 25. Figure~\ref{fig:coshEvolveb} shows the time-evolution simulated by the full circuit compared to the classical simulation for a Trotter time-step of $\delta t = 0.1$. Good agreement is achieved between the quantum and classical simulations, with the KL divergence of the full circuit matching exactly with the compact simulation using the {\tt Ket} command, as shown by the exact match between the evolution using the full circuit and the {\tt Ket} operation in Figure~\ref{fig:fullvsket}. This agreement therefore validates the performance of the method.


\section{Towards quantum field theory}\label{sec:qft}

Given the ability to perform real-time dynamics on a single quantum-mechanical state, continuous-variable models 
of quantum computing open up interesting avenues to explore from the perspective of field theory. Both fundamental and effective quantum fields are of paramount importance in many aspects of physics, in particular in particle physics and the Standard Model. 
Many phenomena, such as strong coupling effects in gauge theory, quantum tunnelling, phase-transitions and other dynamical processes are very hard to study analytically and quantum computing promises to become an important tool, as proposed in the work of References~\cite{Jordan:2011ci,Jordan:2012xnu,Jordan:2014tma,Jordan_2018,Abel:2020qzm,Abel:2020ebj,PhysRevA.105.012412} (see References~\cite{Klco:2018zqz} for a more recent review). 

The reason that the continuous-variable method of quantum computing is an attractive platform for such studies is that fields can be 
encoded without need for explicitly digitising the field value itself. One can instead simply use the continuous variables to stand for field values. This will turn out to be a great simplification because it then allows the kinetic terms in the field theory Hamiltonian to be constructed using a much smaller number of simple Gaussian gates. 

This Section will demonstrate that this can be implemented by outlining a framework for real scalar field theory in 1+1 dimensions. The time dimension will as for the quantum-mechanical system be encapsulated by the Trotterised evolution. This is to be accompanied by a single space dimension which is discretised in $M$ qumodes. The expectation values of the fields $\langle \varphi\rangle $ at each point in space will be encoded in the value of the $\langle \hat x\rangle $ value on each qumode. In order to avoid confusion the single physical space-dimension will be denoted $r$, and it will be discretised using a one-dimensional lattice of spacing $a$. Thus the field at the \nth{k} space position,
\beq 
r_k~=~ r_0 + k \, a ~~~;~~~~k=1\ldots M~,
\eeq 
where $r_0$ is a constant fiducial value,
 is described by the \nth{k} qumode: 
\beq 
\varphi (r_k) ~=~ \hat x_k~.
\label{eq:phi-x}
\eeq 
For a space interval $ r\in [-L/2,L/2]$ we have $a= L/M$. 

In order to set-up the system one may use the fact that the canonical momenta $\hat p_k$ are already included among the available continuous variables (therefore it is not necessary  to implement a matrix representation of the action of $\hat p^2$ in the $\hat x$-basis as one would have to do in the Jordan-Lee-Preskill field discretization method for example \cite{Jordan:2011ci,Jordan:2012xnu,Klco:2018zqz} or in the domain-wall encoding of References~\cite{Abel:2020ebj,Abel:2020qzm}). 
Thus the Hamiltonian discretised over the $M$ space points becomes 
\beq 
\label{eq:qft}
{\cal H} a^{-1} ~=~ \sum_{k=1}^M \left( \frac{1}{2}  \pi ^2_k + \frac{1}{2} (\partial_r  \varphi_k) ^2  + V(\varphi_k)\right) ~.
\eeq
In photonic systems the continuous variable $\hat x$ on a qumode and its conjugate variable $\hat p$ are already canonically normalised, with the commutation relation between the qumodes being 
\beq 
[ \hat x_k, ~\hat p_m   ] ~=~ i \delta_{km}~.
\eeq
However in the discretised field theory the field theoretic conjugate momenta are required to satisfy $[ \varphi(r_k),\pi (r_\ell)]] =i a^{-1}\delta_{k\ell}  $. Therefore  the correct commutation relations for the field and its conjugate momentum are given by identifying
\beq 
\pi(r_k) ~=~ a^{-1} \hat p_k~.
\label{eq:pi-p}
\eeq
Finally the spatial derivative $\partial_r  \phi_k$ can be approximated by using the discretised derivative: 
\begin{align}
 (\partial_r  \varphi_k) ^2 (r) ~&=~  \,  \frac{({\varphi (r_k+a)-\varphi (r_k)} )^2}{a^2}\nonumber \\
 ~&\equiv ~  \,  \frac{({\hat x_{k+1}-\hat x_k} )^2}{a^2} ~.
 \end{align}
 The space-discretised field theory in Equation~\eqref{eq:qft} in terms of the sum over qumode operators then becomes
\begin{align}
{\cal H} a ~&=~ \sum_{k=1}^M \left( \frac{1}{2}  \hat p^2_k + \frac{1}{2} ({\hat x_{\overline{k+1}}-\hat x_k} )^2  + a^2 V(\hat x_k)\right)~,
\label{eq:qumodes-qft}
\end{align}
where it is convenient to adapt periodic coordinates for the  space dimension, such that 
\beq
\overline {k+1} ~=~ k+1 ~~\mbox{ mod($M$)} ~.
\eeq 
The Hamiltonian takes a more familiar form if one expands the terms in the Hamiltonian:
\begin{align}
{\cal H} a ~&=~ \sum_{k=1}^M \left( \frac{1}{2}  \hat p^2_k +\frac{1}{2} {\hat x^2_k} 
 + H_1(\hat x_k)\right) - \sum_{k=1}^M {\hat x_{\overline{k+1}}\hat x_k}   ~,
\end{align}
where again $H_1$ plays the role of an effective potential 
\beq 
H_1(\hat x) ~=~ \frac{1}{2} \hat x^2 + a^2 V (\hat x)  ~.
\eeq
Finally  the overall factor of $a$ may be absorbed by rescaling the evolved time, $\delta t ' = \delta t/a $.

\begin{figure}[t!]
\centering
\includegraphics[width=0.8\textwidth]{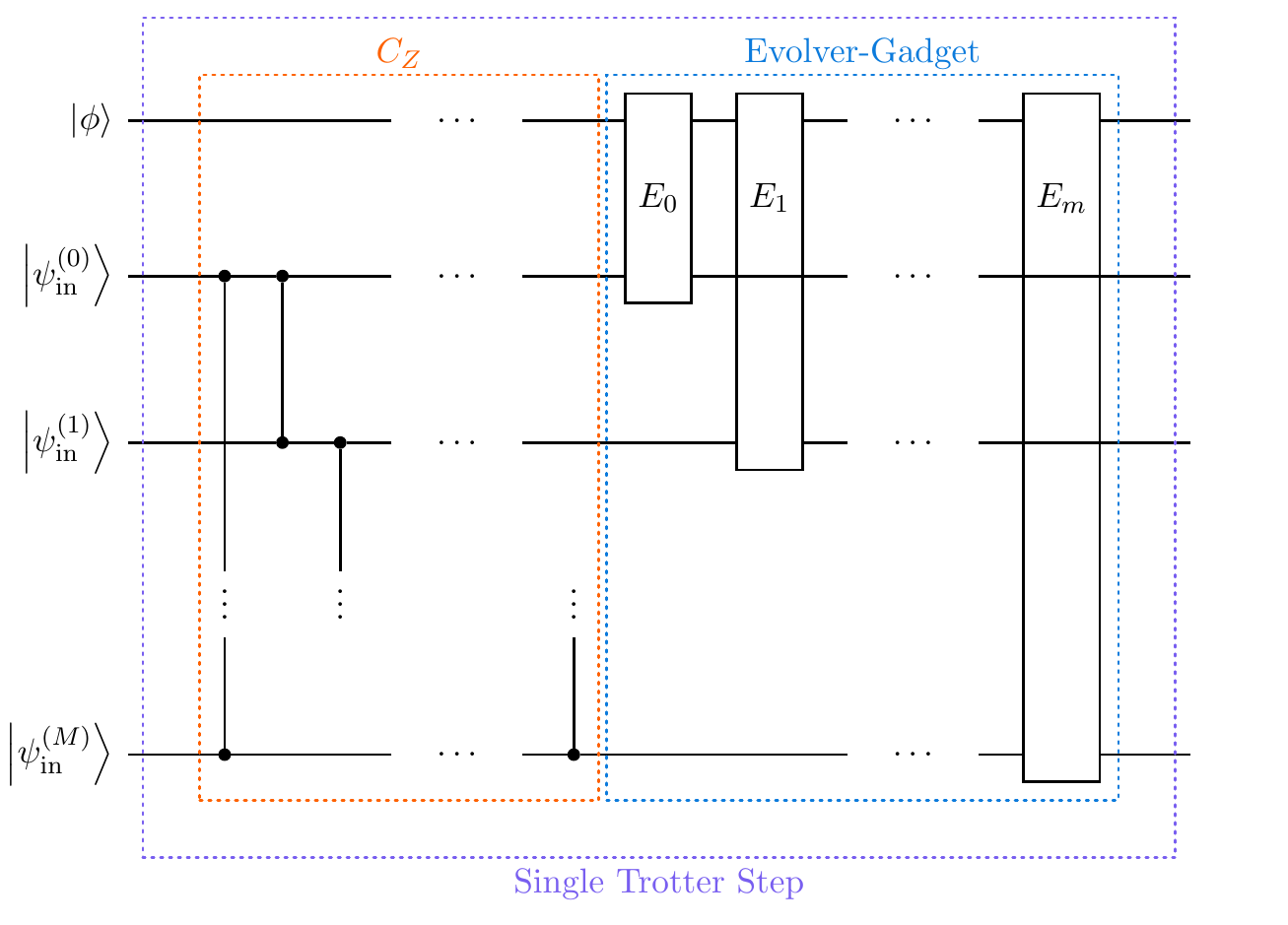}
\caption{Circuit for scalar quantum field theory on $M$ space points.}\label{fig:QFTC}
\end{figure}

The simplicity of qumode implementation is  at this point  notable: the Hamiltonian ultimately consists of a simple sum over terms that exactly resemble the quantum mechanics evolution on each qumode, together with just a single ring of ``hopping terms'' which connect each qumode to its neighbour. These terms are nothing other than controlled-Z gates $ C_Z (\delta t'; x_{\overline{k+1}},x_k ) $.
The entire circuit is shown in Figure~\ref{fig:QFTC}, where the evolver-gadgets, labelled $E_i$, each comprise the circuit shown in Figure~\ref{fig:gadget}.

It is worth comparing the scaling of this method with that of a discrete system in terms of the required gate operations. Each evolver-gadget contains 3 gates. In addition there are $M$ of the $C_Z$ gates. The ancilla circuit for the evolver-state can be reused so this does not need to be included in the circuit count. Thus in total there are $M$ qumodes and  $4M$ gates. In a  $d$-dimensional system this scales as $(4M)^d$ gates. By contrast suppose the field is encoded in  a discrete way, with each field value being encoded by $N$ qubits. To make the kinetic cross-terms, every qubit describing the field at a given space point has to be connected to every qubit of the field at the two-dimensional nearest-neighbour points. Thus one requires at least $M^d \times N^{2d}$ gates, even before the potential has been encoded. As is evident it is the gate-count that gets out-of-hand very quickly. Indeed a three-dimensional lattice that is only 10 points on a side with the field encoded in 10 qubits, which gives only 1/32 accuracy assuming a binary encoding of complex values, requires at least a billion gates\footnote{One might suppose that a momentum basis for the embedding could be beneficial, but then the potential would be even more problematic.}. The same system encoded on a photonic device, including the potential, would require only $40^3=64,000$  gates.


\section{Conclusion} 

We focused on Continuous-Variable Quantum Computing (CVQC) and its applications in simulating quantum mechanics and quantum field theory. Our investigation stems from recognising the need to surpass classical computational paradigms to deepen our understanding of fundamental physics through quantum-mechanical simulations.
Our primary objective was to demonstrate the efficacy of CVQC, leveraging the infinite-dimensional Hilbert space of quantum states, for the accurate simulation of quantum mechanics. We achieved this by meticulously constructing a framework to simulate the time evolution of quantum states under arbitrary Hamiltonians using photonic devices. This involved a detailed exploration of Gaussian and non-Gaussian gate operations essential for the manipulation of quantum states encoded in the continuous observables of photons.
A pivotal technical achievement in our paper is the development of the {\it evolver-state}, a specially prepared quantum state that facilitates the desired Trotterised time-evolution of a quantum-mechanical wavefunction. This approach allowed us to simulate the time evolution of quantum systems under arbitrary potentials, using a combination of quantum gate operations and the strategic manipulation of the evolver-state.

The proposed algorithm for simulating the time-evolution of a quantum-mechanical system under the influence of an arbitrary Hamiltonian has been validated against an exact, classical simulation. The circuit shows good agreement with the classical approach for three scenarios, and has been shown to work in full up to a Fock truncation of $n_{\rm max}=25$, limited by memory constraints when simulating the quantum device. In a practical application on a real CVQC device, these limitations would not be present and the full circuit could be achieved. The promising agreement between these simulations underscores the potential of our approach in simulating complex quantum systems.

Furthermore, we ventured into the domain of quantum field theory, proposing a scheme to discretise space without the need to discretise the field values themselves, thus maintaining the continuous nature of the fields. This proposition opens new avenues for applying CVQC to quantum field theories, potentially simplifying the implementation of these theories on photonic quantum computers.

Thus, this marks a significant stride towards harnessing the capabilities of photonic quantum computing for the simulation of quantum mechanics and the exploration of quantum field theory. We anticipate that our findings will enrich the field of quantum computing for field theories and catalyse further research into the simulation of quantum phenomena.

\newpage
\noindent{\textit{\textbf{Acknowledgments} We thank Christopher Brown for valuable discussions. We acknowledge the use of Xanadu services for this work, and appreciate the support provided through the Xanadu forum.}

\bibliographystyle{inspire}
\bibliography{refs}{}

\providecommand{\href}[2]{#2}\begingroup\raggedright\begin{thebibliography}{10}

\bibitem{Lloyd1999}
S.~Lloyd and S.~L. Braunstein, ``Quantum computation over continuous
  variables,'' \href{http://dx.doi.org/10.1103/PhysRevLett.82.1784}{Phys. Rev.
  Lett. {\bfseries 82} (2, 1999) 1784--1787}.
  \url{https://link.aps.org/doi/10.1103/PhysRevLett.82.1784}.

\bibitem{RevModPhys.77.513}
S.~L. Braunstein and P.~van Loock, ``Quantum information with continuous
  variables,'' \href{http://dx.doi.org/10.1103/RevModPhys.77.513}{Rev. Mod.
  Phys. {\bfseries 77} (Jun, 2005) 513--577}.
  \url{https://link.aps.org/doi/10.1103/RevModPhys.77.513}.

\bibitem{Ladd2010}
T.~D. Ladd, F.~Jelezko, R.~Laflamme, Y.~Nakamura, C.~Monroe, and J.~L. O'Brien,
  ``Quantum computers,'' \href{http://dx.doi.org/10.1038/nature08812}{Nature
  {\bfseries 464} (2010) 45--53}.

\bibitem{Adesso_2014}
G.~Adesso, S.~Ragy, and A.~R. Lee, ``Continuous variable quantum information:
  Gaussian states and beyond,''
  \href{http://dx.doi.org/10.1142/S1230161214400010}{Open Systems \&
  Information Dynamics {\bfseries 21} no.~01n02, (2014) 1440001}.

\bibitem{PhysRevA.73.032318}
J.~Zhang and S.~L. Braunstein, ``Continuous-variable gaussian analog of cluster
  states,'' \href{http://dx.doi.org/10.1103/PhysRevA.73.032318}{Phys. Rev. A
  {\bfseries 73} (Mar, 2006) 032318}.
  \url{https://link.aps.org/doi/10.1103/PhysRevA.73.032318}.

\bibitem{Kok2007}
P.~Kok, W.~J. Munro, K.~Nemoto, T.~C. Ralph, J.~P. Dowling, and G.~J. Milburn,
  ``Linear optical quantum computing with photonic qubits,''
  \href{http://dx.doi.org/10.1103/RevModPhys.79.135}{Reviews of Modern Physics
  {\bfseries 79} (2007) 135--174}.

\bibitem{Slussarenko2019}
S.~Slussarenko and G.~J. Pryde, ``Photonic quantum information processing: A
  concise review,'' \href{http://dx.doi.org/10.1063/1.5115814}{Applied Physics
  Reviews {\bfseries 6} (3, 2019) 41303}.
  \url{https://doi.org/10.1063/1.5115814}.

\bibitem{Killoran_2019}
N.~Killoran, J.~Izaac, N.~Quesada, V.~Bergholm, M.~Amy, and C.~Weedbrook,
  ``Strawberry fields: A software platform for photonic quantum computing,''
  \href{http://dx.doi.org/10.22331/q-2019-03-11-129}{Quantum {\bfseries 3}
  (Mar., 2019) 129}. \url{http://dx.doi.org/10.22331/q-2019-03-11-129}.

\bibitem{Bromley_2020}
T.~R. Bromley, J.~M. Arrazola, S.~Jahangiri, J.~Izaac, N.~Quesada, A.~D. Gran,
  M.~Schuld, J.~Swinarton, Z.~Zabaneh, and N.~Killoran, ``Applications of
  near-term photonic quantum computers: software and algorithms,''
  \href{http://dx.doi.org/10.1088/2058-9565/ab8504}{Quantum Science and
  Technology {\bfseries 5} no.~3, (May, 2020) 034010}.
  \url{http://dx.doi.org/10.1088/2058-9565/ab8504}.

\bibitem{Eaton_2022}
M.~Eaton, A.~Hossameldin, R.~J. Birrittella, P.~M. Alsing, C.~C. Gerry,
  H.~Dong, C.~Cuevas, and O.~Pfister, ``Resolution of 100 photons and quantum
  generation of unbiased random numbers,''
  \href{http://dx.doi.org/10.1038/s41566-022-01105-9}{Nature Photonics
  {\bfseries 17} no.~1, (Dec., 2022) 106–111}.
  \url{http://dx.doi.org/10.1038/s41566-022-01105-9}.

\bibitem{Taballione2023modeuniversal}
C.~Taballione, M.~C. Anguita, M.~de~Goede, P.~Venderbosch, B.~Kassenberg,
  H.~Snijders, N.~Kannan, W.~L. Vleeshouwers, D.~Smith, J.~P. Epping,
  R.~van~der Meer, P.~W.~H. Pinkse, H.~van~den Vlekkert, and J.~J. Renema,
  ``20-{M}ode {U}niversal {Q}uantum {P}hotonic {P}rocessor,''
  \href{http://dx.doi.org/10.22331/q-2023-08-01-1071}{{Quantum} {\bfseries 7}
  (Aug., 2023) 1071}. \url{https://doi.org/10.22331/q-2023-08-01-1071}.

\bibitem{PhysRevA.100.012326}
K.~K. Sabapathy, H.~Qi, J.~Izaac, and C.~Weedbrook, ``Production of photonic
  universal quantum gates enhanced by machine learning,''
  \href{http://dx.doi.org/10.1103/PhysRevA.100.012326}{Phys. Rev. A {\bfseries
  100} (Jul, 2019) 012326}.
  \url{https://link.aps.org/doi/10.1103/PhysRevA.100.012326}.

\bibitem{PhysRevA.100.052301}
D.~Su, C.~R. Myers, and K.~K. Sabapathy, ``Conversion of gaussian states to
  non-gaussian states using photon-number-resolving detectors,''
  \href{http://dx.doi.org/10.1103/PhysRevA.100.052301}{Phys. Rev. A {\bfseries
  100} (Nov, 2019) 052301}.
  \url{https://link.aps.org/doi/10.1103/PhysRevA.100.052301}.

\bibitem{PhysRevLett.86.5188}
R.~Raussendorf and H.~J. Briegel, ``A one-way quantum computer,''
  \href{http://dx.doi.org/10.1103/PhysRevLett.86.5188}{Phys. Rev. Lett.
  {\bfseries 86} (May, 2001) 5188--5191}.
  \url{https://link.aps.org/doi/10.1103/PhysRevLett.86.5188}.

\bibitem{Booth:2021hvw}
R.~I. Booth and D.~Markham, ``{Flow conditions for continuous variable
  measurement-based quantum computing},''
  \href{http://dx.doi.org/10.22331/q-2023-10-19-1146}{Quantum {\bfseries 7}
  (2023) 1146}, \href{http://arxiv.org/abs/2104.00572}{[arXiv:2104.00572
  [quant-ph]]}.

\bibitem{Jordan:2011ci}
S.~P. Jordan, K.~S.~M. Lee, and J.~Preskill, ``{Quantum Computation of
  Scattering in Scalar Quantum Field Theories},'' Quant. Inf. Comput.
  {\bfseries 14} (2014) 1014--1080,
  \href{http://arxiv.org/abs/1112.4833}{[arXiv:1112.4833 [hep-th]]}.

\bibitem{Jordan:2012xnu}
S.~P. Jordan, K.~S.~M. Lee, and J.~Preskill, ``{Quantum Algorithms for Quantum
  Field Theories},'' \href{http://dx.doi.org/10.1126/science.1217069}{Science
  {\bfseries 336} (2012) 1130--1133},
  \href{http://arxiv.org/abs/1111.3633}{[arXiv:1111.3633 [quant-ph]]}.

\bibitem{Jordan:2014tma}
S.~P. Jordan, K.~S.~M. Lee, and J.~Preskill, ``{Quantum Algorithms for
  Fermionic Quantum Field Theories},''
  \href{http://arxiv.org/abs/1404.7115}{[arXiv:1404.7115 [hep-th]]}.

\bibitem{PhysRevA.92.063825}
K.~Marshall, R.~Pooser, G.~Siopsis, and C.~Weedbrook, ``Quantum simulation of
  quantum field theory using continuous variables,''
  \href{http://dx.doi.org/10.1103/PhysRevA.92.063825}{Phys. Rev. A {\bfseries
  92} (Dec, 2015) 063825}.
  \url{https://link.aps.org/doi/10.1103/PhysRevA.92.063825}.

\bibitem{Jordan_2018}
S.~P. Jordan, H.~Krovi, K.~S.~M. Lee, and J.~Preskill, ``Bqp-completeness of
  scattering in scalar quantum field theory,''
  \href{http://dx.doi.org/10.22331/q-2018-01-08-44}{Quantum {\bfseries 2}
  (Jan., 2018) 44}. \url{http://dx.doi.org/10.22331/q-2018-01-08-44}.

\bibitem{Klco:2018zqz}
N.~Klco and M.~J. Savage, ``{Digitization of scalar fields for quantum
  computing},'' \href{http://dx.doi.org/10.1103/PhysRevA.99.052335}{Phys. Rev.
  A {\bfseries 99} no.~5, (2019) 052335},
  \href{http://arxiv.org/abs/1808.10378}{[arXiv:1808.10378 [quant-ph]]}.

\bibitem{Banuls:2019bmf}
M.~C. Ba\~nuls {\em et~al.}, ``{Simulating Lattice Gauge Theories within
  Quantum Technologies},''
  \href{http://dx.doi.org/10.1140/epjd/e2020-100571-8}{Eur. Phys. J. D
  {\bfseries 74} no.~8, (2020) 165},
  \href{http://arxiv.org/abs/1911.00003}{[arXiv:1911.00003 [quant-ph]]}.

\bibitem{Abel:2020qzm}
S.~Abel and M.~Spannowsky, ``{Observing the fate of the false vacuum with a
  quantum laboratory},''
  \href{http://dx.doi.org/10.1103/PRXQuantum.2.010349}{P. R. X. Quantum.
  {\bfseries 2} (2021) 010349},
  \href{http://arxiv.org/abs/2006.06003}{[arXiv:2006.06003 [hep-th]]}.

\bibitem{Abel:2020ebj}
S.~Abel, N.~Chancellor, and M.~Spannowsky, ``{Quantum computing for quantum
  tunneling},'' \href{http://dx.doi.org/10.1103/PhysRevD.103.016008}{Phys. Rev.
  D {\bfseries 103} no.~1, (2021) 016008},
  \href{http://arxiv.org/abs/2003.07374}{[arXiv:2003.07374 [hep-ph]]}.

\bibitem{PRXQuantum.4.027001}
C.~W. Bauer, Z.~Davoudi, A.~B. Balantekin, T.~Bhattacharya, M.~Carena, W.~A.
  de~Jong, P.~Draper, A.~El-Khadra, N.~Gemelke, M.~Hanada, D.~Kharzeev,
  H.~Lamm, Y.-Y. Li, J.~Liu, M.~Lukin, Y.~Meurice, C.~Monroe, B.~Nachman,
  G.~Pagano, J.~Preskill, E.~Rinaldi, A.~Roggero, D.~I. Santiago, M.~J. Savage,
  I.~Siddiqi, G.~Siopsis, D.~Van~Zanten, N.~Wiebe, Y.~Yamauchi,
  K.~Yeter-Aydeniz, and S.~Zorzetti, ``Quantum simulation for high-energy
  physics,'' \href{http://dx.doi.org/10.1103/PRXQuantum.4.027001}{PRX Quantum
  {\bfseries 4} (May, 2023) 027001}.
  \url{https://link.aps.org/doi/10.1103/PRXQuantum.4.027001}.

\bibitem{Kane:2022ejm}
C.~Kane, D.~M. Grabowska, B.~Nachman, and C.~W. Bauer, ``{Efficient quantum
  implementation of 2+1 U(1) lattice gauge theories with Gauss law
  constraints},'' \href{http://arxiv.org/abs/2211.10497}{[arXiv:2211.10497
  [quant-ph]]}.

\bibitem{Blance:2020ktp}
A.~Blance and M.~Spannowsky, ``{Unsupervised event classification with graphs
  on classical and photonic quantum computers},''
  \href{http://dx.doi.org/10.1007/JHEP08(2021)170}{JHEP {\bfseries 21} (2020)
  170}, \href{http://arxiv.org/abs/2103.03897}{[arXiv:2103.03897 [hep-ph]]}.

\bibitem{Bepari:2020xqi}
K.~Bepari, S.~Malik, M.~Spannowsky, and S.~Williams, ``{Towards a quantum
  computing algorithm for helicity amplitudes and parton showers},''
  \href{http://dx.doi.org/10.1103/PhysRevD.103.076020}{Phys. Rev. D {\bfseries
  103} no.~7, (2021) 076020},
  \href{http://arxiv.org/abs/2010.00046}{[arXiv:2010.00046 [hep-ph]]}.

\bibitem{Bepari:2021kwv}
K.~Bepari, S.~Malik, M.~Spannowsky, and S.~Williams, ``{Quantum walk approach
  to simulating parton showers},''
  \href{http://dx.doi.org/10.1103/PhysRevD.106.056002}{Phys. Rev. D {\bfseries
  106} no.~5, (2022) 056002},
  \href{http://arxiv.org/abs/2109.13975}{[arXiv:2109.13975 [hep-ph]]}.

\bibitem{Gustafson:2022dsq}
G.~Gustafson, S.~Prestel, M.~Spannowsky, and S.~Williams, ``{Collider events on
  a quantum computer},'' \href{http://dx.doi.org/10.1007/JHEP11(2022)035}{JHEP
  {\bfseries 11} (2022) 035},
  \href{http://arxiv.org/abs/2207.10694}{[arXiv:2207.10694 [hep-ph]]}.

\bibitem{Schuld2015}
I.~S. Maria~Schuld and F.~Petruccione, ``An introduction to quantum machine
  learning,''
  \href{http://dx.doi.org/10.1080/00107514.2014.964942}{Contemporary Physics
  {\bfseries 56} no.~2, (2015) 172--185}.
  \url{https://doi.org/10.1080/00107514.2014.964942}.

\bibitem{Carleo:2019ptp}
G.~Carleo, I.~Cirac, K.~Cranmer, L.~Daudet, M.~Schuld, N.~Tishby,
  L.~Vogt-Maranto, and L.~Zdeborov\'a, ``{Machine learning and the physical
  sciences},'' \href{http://dx.doi.org/10.1103/RevModPhys.91.045002}{Rev. Mod.
  Phys. {\bfseries 91} no.~4, (2019) 045002},
  \href{http://arxiv.org/abs/1903.10563}{[arXiv:1903.10563 [physics.comp-ph]]}.

\bibitem{Abel:2022lqr}
S.~Abel, J.~C. Criado, and M.~Spannowsky, ``{Completely quantum neural
  networks},'' \href{http://dx.doi.org/10.1103/PhysRevA.106.022601}{Phys. Rev.
  A {\bfseries 106} no.~2, (2022) 022601},
  \href{http://arxiv.org/abs/2202.11727}{[arXiv:2202.11727 [quant-ph]]}.

\bibitem{Abel:2023erv}
S.~Abel, J.~C. Criado, and M.~Spannowsky, ``{Training Neural Networks with
  Universal Adiabatic Quantum Computing},''
  \href{http://arxiv.org/abs/2308.13028}{[arXiv:2308.13028 [quant-ph]]}.

\bibitem{Rousselot:2023pcj}
A.~Rousselot and M.~Spannowsky, ``{Generative Invertible Quantum Neural
  Networks},'' \href{http://arxiv.org/abs/2302.12906}{[arXiv:2302.12906
  [hep-ph]]}.

\bibitem{PhysRevA.97.062311}
T.~Kalajdzievski, C.~Weedbrook, and P.~Rebentrost, ``Continuous-variable gate
  decomposition for the bose-hubbard model,''
  \href{http://dx.doi.org/10.1103/PhysRevA.97.062311}{Phys. Rev. A {\bfseries
  97} (Jun, 2018) 062311}.
  \url{https://link.aps.org/doi/10.1103/PhysRevA.97.062311}.

\bibitem{PhysRevA.105.012412}
K.~Yeter-Aydeniz, E.~Moschandreou, and G.~Siopsis, ``Quantum imaginary-time
  evolution algorithm for quantum field theories with continuous variables,''
  \href{http://dx.doi.org/10.1103/PhysRevA.105.012412}{Phys. Rev. A {\bfseries
  105} (Jan, 2022) 012412}.
  \url{https://link.aps.org/doi/10.1103/PhysRevA.105.012412}.

\bibitem{EISERT2003}
J.~Eisert and M.~B. Plenio, ``Introduction to the basics of entanglement theory
  in continuous-variable systems,''
  \href{http://dx.doi.org/10.1142/S0219749903000371}{International Journal of
  Quantum Information {\bfseries 01} (2003) 479--506}.
  \url{https://doi.org/10.1142/S0219749903000371}.

\bibitem{Hamilton2017}
C.~S. Hamilton, R.~Kruse, L.~Sansoni, S.~Barkhofen, C.~Silberhorn, and I.~Jex,
  ``Gaussian boson sampling,''
  \href{http://dx.doi.org/10.1103/PhysRevLett.119.170501}{Phys. Rev. Lett.
  {\bfseries 119} (10, 2017) 170501}.
  \url{https://link.aps.org/doi/10.1103/PhysRevLett.119.170501}.

\bibitem{Clements2016}
W.~R. Clements, P.~C. Humphreys, B.~J. Metcalf, W.~S. Kolthammer, and I.~A.
  Walmsley, ``Optimal design for universal multiport interferometers,''
  \href{http://dx.doi.org/10.1364/OPTICA.3.001460}{Optica {\bfseries 3} (12,
  2016) 1460--1465}.
  \url{https://opg.optica.org/optica/abstract.cfm?URI=optica-3-12-1460}.

\bibitem{PhysRevLett.117.110801}
H.~Vahlbruch, M.~Mehmet, K.~Danzmann, and R.~Schnabel, ``Detection of 15 db
  squeezed states of light and their application for the absolute calibration
  of photoelectric quantum efficiency,''
  \href{http://dx.doi.org/10.1103/PhysRevLett.117.110801}{Phys. Rev. Lett.
  {\bfseries 117} (Sep, 2016) 110801}.
  \url{https://link.aps.org/doi/10.1103/PhysRevLett.117.110801}.

\bibitem{qibo_paper}
S.~Efthymiou, S.~Ramos-Calderer, C.~Bravo-Prieto, A.~P{\'{e}}rez-Salinas,
  D.~Garc{\'{\i}}a-Mart{\'{\i}}n, A.~Garcia-Saez, J.~I. Latorre, and
  S.~Carrazza, ``Qibo: a framework for quantum simulation with hardware
  acceleration,'' \href{http://dx.doi.org/10.1088/2058-9565/ac39f5}{Quantum
  Science and Technology {\bfseries 7} no.~1, (Dec, 2021) 015018}.
  \url{https://doi.org/10.1088/2058-9565/ac39f5}.

\bibitem{kullback1951information}
S.~Kullback and R.~A. Leibler, ``On information and sufficiency,'' The annals
  of mathematical statistics {\bfseries 22} no.~1, (1951) 79--86.

\end{thebibliography}\endgroup

\end{document}